# Natural fumigation as a mechanism for volatile transport between flower organs

Benoît Boachon[1†], Joseph H. Lynch[1,5], Shaunak Ray[2], Jing Yuan[3,5], Kristian Mark P. Caldo[1], Robert R. Junker[4], Sharon A. Kessler[3,5], John A. Morgan[1,2] & Natalia Dudareva[1,5*].

[1]Department of Biochemistry, Purdue University, West Lafayette, Indiana, U.S.A.

[2]Davidson School of Chemical Engineering, Purdue University, West Lafayette, Indiana, U.S.A.

[3]Department of Botany and Plant Pathology, Purdue University, West Lafayette, Indiana, U.S.A.

[4]Department of Biosciences, University Salzburg, Salzburg, Austria.

[5]Purdue Center for Plant Biology, Purdue University, West Lafayette, Indiana, U.S.A.

[†]Present address: BVpam FRE 3727, Univ Lyon-UJM-Saint-Etienne-CNRS, Saint-Etienne, France.

*Corresponding author: dudareva@purdue.edu

**Editorial Summary:** Bioactive sesquiterpenes accumulating in petunia stigmas are synthesized in the floral tube, and are then transported to the pistil via natural fumigation within the internal airspace of the developing flower.

## Abstract

Plants synthesize volatile organic compounds (VOCs) to attract pollinators and beneficial microorganisms, to defend themselves against herbivores and pathogens and for plant-plant communication. Generally, accumulation and emission of VOCs occur from the tissue of their biosynthesis. However, using biochemical and reverse genetic approaches, we demonstrate a new physiological phenomenon: inter-organ aerial transport of VOCs via natural fumigation. Before petunia flowers open, a tube-specific terpene synthase produces sesquiterpenes, which are released inside the buds and then accumulate in the stigma, potentially defending the developing stigma from pathogens. These VOCs also affect reproductive organ development and seed yield, which is a previously unknown function for terpenoid compounds.

## Main
## Introduction

Decades of research have uncovered essential roles of volatile organic compounds (VOCs) in plant fitness. VOCs are lipophilic, low molecular weight compounds with high vapor pressure at ambient temperatures, which have been shown to be instrumental in plant-plant, plant-animal, and plant-microbe interactions[1,2]. Chemically diverse, VOCs are mainly classified as terpenoids, phenylpropanoids/benzenoids, and fatty acid and amino acid derivatives. They are typically released from the tissue(s) of origin directly into the surrounding environment, usually with low accumulation within those tissues. Larger pools of VOCs may also be sequestered in the tissue of biosynthesis, either physically in specialized structures like trichomes[3], or chemically via VOC modifications that prevent their volatilization[4,5]. VOCs have been shown to be an essential part of the plant defense response, including plant priming, in which attacked plants emit compounds that are perceived by leaves of neighboring plants without detectable accumulation within the receptive tissues[6]. Although volatile signaling between leaves of the same or nearby plants has been reported, the involvement of VOCs in inter-organ communication remains largely unexplored.

Of all plant organs, flowers emit the highest levels of VOCs. Many of these VOCs are used to either attract specific pollinators or repel antagonist insects[7–9], thus ensuring plant reproductive and evolutionary success. Flowers are also an ideal habitat for and highly susceptible to pathogens and florivores due to their high nutrient content and low lignification of cell walls that serve as a physical barrier against penetration. Within the flower, stigmas are especially susceptible to pathogens. Their moist and nutritive environment, which promotes germination and growth of pollen grains, can also facilitate growth of microorganisms[10]. However, the molecular mechanisms responsible for protection of reproductive organs, and subsequent overall plant survival in natural ecosystems, still remain to be determined.

*Petunia hybrida* flowers are an excellent model system for investigating the biosynthesis, regulation, and emission of plant VOCs. Although petunia flowers produce predominantly phenylpropanoid/benzenoid compounds, low levels of two sesquiterpene compounds, germacrene D (**1**) and β-cadinene (**2**) (cadina-3,9-diene), have also been detected in *in vitro* headspace of CaCl$_2$ extracts[11]. In an attempt to understand the biosynthesis of these terpenoid volatiles and their role(s) in petunia flowers, we identified and characterized terpene synthases (TPSs) responsible for terpenoid production. Moreover, we discovered natural fumigation (i.e. gas treatment of an enclosed space) as a mechanism for directional inter-organ terpenoid transport, and a previously unidentified function for volatile terpenoids in development of reproductive organs.

## Results

### Petunia TPSs and fate of their products

Using targeted metabolite profiling of volatiles present in petunia floral organs we detected two additional sesquiterpenes, bicyclogermacrene (**3**) and germacrene D-4-ol (**4**), and the monoterpene geraniol (**5**) along with previously detected terpenes, germacrene D and β-cadinene (Supplementary Fig. 1). The highest levels of sesquiterpenes were found in pistils, with significantly lower levels in other floral organs including stamens, tubes and corollas (Supplementary Fig. 1). No sesquiterpene glucosides were detected in any floral tissue. In contrast, geraniol and its glucoside were found only in pistils and were absent in other analyzed tissues (Supplementary Figs. 1 and 2a). In addition, terpene levels did not display any time-dependent changes throughout the day (Supplementary Fig. 2b).

To identify genes encoding enzymes responsible for terpene production, we searched petunia flower RNA-seq data sets[12] for terpene synthase (TPS) genes. This search resulted in four putative TPSs designated *PhTPS1* to *PhTPS4* numbered according to their expression levels (Supplementary Fig. 3). Phylogenetic analysis revealed that PhTPS1, PhTPS3 and PhTPS4 all belong to the TPS-a clade, which contains most of the characterized sesquiterpene synthases[13], whereas PhTPS2 belongs to the TPS-b clade, which includes monoterpene synthases (Supplementary Fig. 4). Indeed, expression of PhTPS2 in *Saccharomyces cerevisiae* resulted in geraniol production (Supplementary Fig. 5a). Furthermore, its transcript levels measured by quantitative RT-PCR (qRT-PCR) were found predominantly in petunia pistils and positively correlated with geraniol accumulation over flower development (Supplementary Fig. 5b and c).

To test whether the petunia sesquiterpene synthase candidates are responsible for biosynthesis of sesquiterpenes detected in flower organs, product specificities of PhTPS1, PhTPS3 and PhTPS4 were determined by analyzing the terpenoid profiles in yeast expressing each of the enzymes (Fig. 1a). PhTPS1 produced a dozen sesquiterpenes with major products being germacrene D, bicyclogermacrene, β-cadinene and germacrene D-4-ol. PhTPS4 generated mainly β-cadinene with a small amount of nerolidol, while PhTPS3 produced only a small amount of β-cadinene (Fig. 1a).

Of sesquiterpenes detected in pistils, β-cadinene was the most abundant compound. Its accumulation, as well as that of other sesquiterpenes, was developmentally regulated with the highest levels on day 2 after floral opening (post-anthesis) (Fig. 1b). Despite the fact that all three petunia sesquiterpene synthases produce β-cadinene (Fig 1a), its accumulation in pistils spatially correlated only with *PhTPS3* and *PhTPS4* expression (Fig. 1b and c). Unexpectedly, germacrene D, bicyclogermacrene and germacrene D-4-ol, the sesquiterpenes produced only by PhTPS1, accumulated mainly in the pistil (Fig. 1b) at ratios similar to their production by recombinant protein and their occurrence in tubes (Supplementary Fig. 6), even though *PhTPS1* expression occurred almost exclusively in the tube (Fig. 1c). Interestingly, *PhTPS1* mRNA was not evenly distributed across the tube. The maximum expression occurred mainly at the top of the tube just below the unexpanded corolla (Supplementary Fig. 7a), the part of the tube in closest proximity to the developing stigma (Supplementary Fig. 7b) where PhTPS1 products accumulate the most (Supplementary Fig. 7c). Moreover, *PhTPS1* expression was developmentally regulated, peaking one day ahead of sesquiterpene levels in pistil (Fig. 1b and Supplementary Fig. 7d), as has been observed for other VOC biosynthetic genes[14].

Consistent with expression data, crude protein extracts prepared from pistils produced only β-cadinene (but not the other sesquiterpenes) when incubated with *e,e*-farnesyl diphosphate (FPP), the sesquiterpene precursor (Supplementary Fig. 8a and b). Similarly, activity for germacrene, bicyclogermacrene and germacrene D-4-ol formation was only detected in the protein crude extracts prepared from the top of the tube (Supplementary Fig. 8c), where expression of *PhTPS1* was the highest (Supplementary Fig. 7a and Fig. 1c).

**Inter-organ transport of terpenes via bud headspace**

The absence of *PhTPS1* expression and TPS activity capable of producing germacrene D, bicyclogermacrene and germacrene D-4-ol in pistils along with simultaneous accumulation of these terpenoids in the stigma suggests that PhTPS1 products detected in the stigma originate in the tube. Thus, we hypothesized that these sesquiterpenes are emitted from the tube into the bud's headspace and accumulate in reproductive organs before flower opening, likely protecting the nutrient-rich pistil from attacking pathogens and florivores via natural fumigation.

To test this aerial transport hypothesis, we first compared the emission rates of PhTPS1 products from the inner (adaxial) and outer (abaxial) surfaces of the petunia tube by using direct-contact sorptive extraction (DCSE) with polydimethylsiloxane coated stir bars (Twisters)[15]. We found that ~ 65 to 75% of each compound was emitted from the inner side of the tube (Supplementary Fig. 9), suggesting a directional release of terpenoids into the headspace of closed buds. Second, tubes were removed from flower buds four days before anthesis, and accumulation of PhTPS1 products was analyzed in pistils at day 1 post-anthesis (Supplementary Fig. 10a). Germacrene D, bicyclogermacrene and germacrene D-4-ol were almost completely absent in pistils from flowers grown without tubes relative to control flowers (Fig. 2a). These pistils, however, still accumulated geraniol and β-cadinene most likely due to the respective activities of PhTPS2 and PhTPS3/PhTPS4 present in this organ (Fig. 2a).

Third, we assessed the movement of PhTPS1 products from the tube to the pistil by feeding only the tubes with a stable isotope-labeled precursor of sesquiterpenes and tracking the fate of labeled products in detached pistils (Supplementary Fig. 10b). Feeding of petunia tubes from day 0 flowers with [2-$^{13}$C]-mevalonolactone, a precursor of FPP, led to accumulation of labeled sesquiterpenes not only in tubes, but also in pistils (Fig. 2b), even though pistils were detached and protected by microvials from contact with [2-$^{13}$C]-mevalonolactone (Supplementary Fig. 10b). These results provide direct evidence for gas phase transmission of stable isotope labelled terpenoids between these flower organs (Fig. 2b).

Finally, we generated transgenic petunia plants with constitutive RNAi downregulation of *PhTPS1* expression (Supplementary Fig. 11). The three independent lines with *PhTPS1* transcript levels reduced by 92-94% barely emitted PhTPS1 products from tubes (Fig. 2c) and had drastically reduced accumulation of respective sesquiterpenes in pistils (Fig. 2d). To directly test for gas phase transport, we performed a complementation experiment in which *PhTPS1*-RNAi pistils were placed within wild-type tubes for 24 h (Supplementary Fig. 10c). Volatiles released from the wild-type tubes were indeed sufficient to restore the accumulation of missing sesquiterpenes in transgenic pistils (Fig. 2e). Analysis of expression of other *PhTPSs* in pistils of transgenic plants revealed an unexpected decrease in *PhTPS4* expression, while *PhTPS2* and *PhTPS3* were unaffected (Supplementary Fig. 12). Although the reasons for this reduction are unknown, it correlated with the observed decrease in β-cadinene levels in transgenic pistils (Fig. 2d), an effect which persisted even upon complementation with wild-type tubes (Fig. 2e).

**Sesquiterpene fumigation alters pistil microbiome**

Flower terpenoids are known to affect the growth of bacterial communities and to protect reproductive organs from microbial pathogens[16], thus we hypothesized that terpene fumigation might affect the bacterial and fungal communities present on the pistil surface. Therefore, we characterized microbiome of wild type and transgenic pistils. Fungal operational taxonomic units (OTUs) were not detected in pistils of either wild-type or *PhTPS1*-RNAi flowers. Bacterial diversity and number of reads were found to be very low (only 7 OTUs) on the surface of pre-anthesis stigma and even less on day 2 post-anthesis (Supplementary Fig. 13a and Supplementary Dataset) demonstrating a low bacterial density on petunia stigmas in general. However, lack of sesquiterpene fumigation on pistils of *PhTPS1*-RNAi flowers resulted in a statistically significant increase in the most abundant bacterial OTU in our samples, belonging to the family Pseudomonas, compared to wild-type pistils (Supplementary Fig. 13b). Furthermore, bacterial community composition responded to time after anthesis and plant genotype (wildtype or transgenic, Supplementary Fig. 13c). These results suggest that fumigation affects bacterial growth on stigmas and also support the role of terpenoids in shaping bacterial communities of petunia reproductive organs[17].

**Floral fumigation affects pistil growth and seed yield**

Even though there is no detectable *PhTPS1* expression in developing pistils, down-regulation of *PhTPS1* transcript levels and the corresponding loss of sesquiterpene fumigation had a striking effect on pistil development. *PhTPS1*-RNAi pistils had significantly lower weights than wild-type pistils (~78 to 86% of wild-type flowers) (Fig 3a and Supplementary Fig. 14a and b), with smaller stigmas (Supplementary Fig. 14d, e and f). Transgenic flowers also had slightly decreased style length with reduced diameter (Supplementary Fig. 14c, e and f). Further experiments showed that pistil size phenotype is independent of pistil genotype and instead depends on tube genotype. *In vitro* growth of *PhTPS1*-RNAi pistils within wild-type tubes (Supplementary Fig. 10d) resulted in recovery of the pistil size phenotypes (Fig. 3b and Supplementary Fig. 14g). This results directly confirmed that PhTPS1-produced volatiles fumigated from the tube regulate pistil growth. Moreover, wild-type pistils grown within *PhTPS1*-RNAi tubes exhibited a similar reduced pistil size phenotype to transgenic pistils (Fig.

3b and Supplementary Fig. 14g), further supporting that terpenoids produced by PhTPS1 in tubes are required for normal pistil development.

Disrupting *PhTPS1* expression also affected seed yield, but not their weight (Fig. 3c). *PhTPS1*-RNAi flowers produced up to 33% fewer seeds than wild-type flowers (Fig. 3c) without any effect on pollen tube growth through the pistils after pollination (Supplementary Fig. 15). These data reveal a role of aerial transport of sesquiterpenes in pistil development and seed yield, and thus successful reproduction (or fitness).

**Pistils contain more cuticular waxes than tubes**

VOCs in general, including sesquiterpenes, are lipophilic and can readily partition from air spaces into the epicuticles of plants[18]. Therefore, the directional transport of the fumigated sesquiterpenes from the tube and their accumulation in the pistil, rather than other tissues of the bud, may be the result of stigma surface physico-chemical properties. Indeed, comparative analysis of waxes revealed that pistils contain a significantly higher level of waxes than tubes (Supplementary Fig. 16a), which could cause the preferential adsorption of sesquiterpenes to the pistil surface. Pistils exposed to artificial fumigation with β-caryophyllene showed time-dependent accumulation (Supplementary Fig. 16b) and accumulated the highest levels of exogenous sesquiterpene over stamen and tube, further demonstrating the pistil's ability to preferentially trap volatiles (Supplementary Fig. 16c).

# Discussion

Volatile terpenoids constitute the majority of plant VOCs and are dominated by mono- and sesquiterpenes. As a part of plant defense, they not only repel and intoxicate attacking herbivores above- and below-ground, and mediate plant-insect and plant-microbial interactions, but are also instrumental in plant-plant communication including priming[1,19,20]. Plant-plant communication via VOCs includes intra- and inter-specific signaling as well as within-plant self-signaling between different branches or adjacent leaves. In all these cases, plant VOCs are first released into the atmosphere to create a diffuse signal that is then perceived by leaves of the same or neighboring plants. While it is clear that receiving plants perceive and react to VOC signal(s), no accumulation of free VOCs has been detected inside recipient tissues. In contrast, in petunia pistils we found high accumulation of sesquiterpenes, which were produced and released by surrounding tubes. This occurs via natural fumigation in the enclosed space of floral buds before flower opening (Fig. 4).

The fumigation of pistils with sesquiterpenes likely allows plants to protect their reproductive organs against microorganisms. For instance, *Arabidopsis thaliana* stigmas produce a sesquiterpene, (*E*)-β-caryophyllene to inhibit the growth of a *Pseudomonas syringae*, a pathogen that causes seed defects[17]. In general, volatile organic compounds have been shown to affect the growth of microorganisms associated with plant tissues[16,21]. Our data suggest that the fumigation of the pistils reduces the density of bacteria colonizing petunia pistils (Supplementary Fig. 13), thereby potentially protecting this tissue against pathogens. Moreover, recent studies demonstrated the effect of floral microbes on pollinator behavior[22,23]. Therefore, mechanisms preventing the colonization of important reproductive tissues by bacteria may be beneficial for plant reproduction even beyond their function in protection against pathogens.

Since the TPS1-produced sesquiterpenes are beneficial to the stigma development, it is not clear why pistils have to be fumigated by tubes rather than synthesizing these compounds *de novo* by themselves as occurs for β-cadinene (Figs. 1a, b and 2a). However, since the absence of tube-produced sesquiterpenes has detrimental effects on pistil development and seed yield (Fig. 3 and Supplementary Fig. 14), it is likely that aerial transmission of volatiles from tubes might serve as a signal, allowing coordination of growth of different floral organs. In outcrossing species such as *Petunia hybrida,* regulation of pistil development by volatile terpenoids from surrounding tube could serve as a mechanism to coordinate the timing of pistil maturation with petal development in order to ensure that the stigma is receptive when the flowers are most likely to attract pollinators. Such "hormone-like" action would be analogous to the well-established ethylene signaling that coordinates senescence of the petal tissue with successful pollination[24,25].

To date, there are a plethora of examples where volatile signals serve as developmental cues. Ethylene, the function of which is not limited to flowers, serves as a key growth regulator impacting leaf, root, shoot and fruit development[26]. Many of these ethylene roles overlap with those of another volatile hormone, methyl jasmonate. Methyl jasmonate, though generally considered a defense compound due to its involvement in responses to biotic and abiotic stresses, has a wide-ranging influence on a variety of developmental processes including seed germination, root growth, fruit ripening and senescence[27,28]. Another example of mobile signal is methyl salicylate, which plays an important role in plant defense[19,29].

Our results showing that volatile sesquiterpenes are necessary for optimal pistil growth (Fig. 3a, b and Supplementary Fig. 14) and represent the first example of plants using their own sesquiterpenes to regulate flower development. This is consistent with previous results showing that terpenoid compounds can potentially influence plant growth. However, terpenoids in these cases were produced by microrhizal fungi[30]. β-Caryophyllene produced by *Fusarium oxysporum* was shown to increase root and shoot length, as well as fresh weight of lettuce (*Lactuca sativa)* seedlings[31]. Lateral root growth-promoting activity was also reported for another sesquiterpene, (-)-thujopsene, released by *Laccaria bicolor*, in grey poplars and Arabidopsis[32].

Flower development has been extensively studied from a molecular genetic perspective[33,34]. However, the function of terpenoids in this process has been overlooked, likely due to their low levels before flower opening. It is possible that the directional inter-organ transport of volatile terpenoids and subsequent hormone-like effects on pistil growth and seed yield observed in this study are linked to floral morphology. Further studies are required to assess whether natural fumigation is conserved in flowering plants, to uncover the mechanisms involved, and to determine its evolutionary advantage in plant reproduction.

**Acknowledgments:** This work was supported by grant IOS-1655438 from the National Science Foundation for JAM and ND and by the USDA National Institute of Food and Agriculture Hatch project 177845 to ND.

**Author contributions:** B.B. and N.D. conceived the study. B.B., S.R., J.Y., J.H.L., R.R.J. S.A.K., J.A.M. and N.D. planned experiments. B.B. performed metabolic profiling, identification and characterization of TPSs, analysis of TPS activities in planta, stable isotope labeling, expression analysis, generation of transgenic plants and their analysis, complementation experiments. S.R. performed wax analysis. J.Y. performed microscopy analysis. K.M.P.C.

analyzed seed yield. J.H.L. performed complementation experiments, expression analysis of TPSs in transgenic plants and metabolic profiling. R.R.J. performed microbiome analysis. B.B., S.R., J.Y., K.M.P.C., J.H.L., R.R.J., S.A.K., J.A.M. and N.D. analyzed and interpreted data. S.A.K., J.A.M. and N.D. supervised the study. B.B., J.H.L. and N.D. wrote the paper. All authors read and edited the manuscript.

**Competing financial interests:** The authors declare no competing financial interests.

# Figure legends

**Fig. 1: Characterization of sesquiterpene synthases expressed in petunia flowers.**

(**a**) Products of PhTPS1, PhTPS3 and PhTPS4 or empty vector (EV) expressed in yeast strain WAT11. GC-MS chromatograms of volatiles emitted from yeast cultures are presented as total extracted ion current (EIC, m/z 93 + 161). Sesquiterpenes were identified by comparison of their mass spectra (MS) to the NIST library. Germacrene D was confirmed by comparison with an authentic standard. Asterisks represent unidentified putative sesquiterpenes. GC-MS chromatograms are representative of three independent experiments. (**b**, **c**) Internal pools of sesquiterpenes (**b**) and relative transcript levels of *PhTPS1*, *PhTPS3* and *PhTPS4* (**c**) in petunia stamen, pistil, tube and corolla over flower development (1-2 cm buds, 3-4 cm buds, flower day 0 before anthesis, and flower day 2 after anthesis) determined by GC-MS (**b**) and by qRT-PCR (**c**), respectively. Internal pools were quantified based on the ratio of the integrated peak area of terpenoids relative to the IS peak area and normalized to the weight of the tissue. *PhPP2A* was used as a reference gene in qRT-PCR. Data are means ± SE (n = 3 biological replicates).

**Fig. 2:| Analysis of inter-organ transport of PhTPS1 products in petunia buds.**

(**a**) Effect of flower tube on accumulation of terpenoids in pistils. Terpenoids were analyzed in pistils on day 1 post-anthesis from intact flowers (control) and from flowers from which tubes were removed 3 days before anthesis. Data are means ± SE (n = 3 biological replicates). (**b**) GC-MS analysis of sesquiterpenes produced by petunia tubes fed with [2-$^{13}$C]-mevalonolactone for 24 h. Amounts of unlabeled and labelled sesquiterpenes were quantified with the specific ions 161 m/z and 162 + 163 m/z (M + 1 and M + 2), respectively. Data are means ± SE (n = 3 biological replicates). (**c** and **d**) Effect of *PhTPS1*-RNAi downregulation on terpenoid emission from tubes over 24 h beginning on day 0 before anthesis (**c**) and their accumulation in pistils on day 1 post-anthesis (**d**). WT, wild type control; EV, empty vector control, and *PhTPS1*-RNAi lines. Data are means ± SE (n = 4 biological replicates). (**e**) Effect of gas phase complementation by wild type tubes on internal pools of mono- and sesquiterpenes in *PhTPS1*-RNAi-11 pistils. Terpenoids were extracted from pistils after 24 h of complementation and analyzed by GC-MS. Data are means ± SE (n = 8 biological replicates). Experimental setups for (**a**), (**b**) and (**e**) are illustrated on Supplementary Fig. 10. Significant p values (p < 0.05), as determined by two-tailed paired t-test relative to control, are shown.

**Fig. 3: Effect of sesquiterpene fumigation on pistil development and seed yield.**

(**a**) Weight of pistils from wild type (WT), an empty vector (EV) control line and *PhTPS1*-RNAi lines on day 1 post-anthesis. Data are means ± SE (n = 30 biological replicates). (**b**) Complementation via gas phase of in vitro growth of *PhTPS1*-RNAi-11 pistils with WT tubes. Experimental setup is illustrated on Supplementary Fig. 10d and described in Materials and Methods. After 4 days of complementation, weight and major and minor axes of stigma were measured. Data are means ± SE (n = 8 biological replicates). (c) Effect of *PhTPS1*-RNAi downregulation on seed production. Total weight of seeds per flower, weight per seed and number of seeds per flower were measured in WT and *PhTPS1*-RNAi transgenic lines. Data are means ± SE (n = 8 biological replicates). Significant p values ($p < 0.05$), as determined by two-tailed paired t-test relative to control, are shown.

## Online Methods

**Plant materials and growth conditions**
*Petunia hybrida* cv. Mitchell diploid (W115; Ball Seed Co., West Chicago, IL) wild-type and *PhTPS1*-RNAi transgenic plants were grown under standard greenhouse conditions as previously described [35]. The *PhTPS1*-RNAi construct was generated by using the Sol Genomics Network VIGS Tool (http://vigs.solgenomics.net/) [36] to identify the best target region of *PhTPS1* coding sequence and verify that the designed *PhTPS1* dsRNA trigger would not result in off-target interference. The *PhTPS1*-RNAi construct included two spliced *PhTPS1* cDNA fragments corresponding to nucleotides 937-1436 and 937-1237 (in antisense orientation) to create a hairpin structure. The sequence was then synthesized by Genscript (Piscataway, NJ) with flanking *AttL1* and *AttL2* sequences for LR recombination (Thermo Fisher Scientific, Waltham, MA, USA) into the pB2WG7 binary vector under the control of the CaMV 35S promoter. *PhTPS1*-RNAi transgenic plants were generated via *Agrobacterium tumefaciens* (strain GV3101) mediated transformation using the standard leaf disk transformation method [37] and 16 transgenic lines were screened for *PhTPS1* mRNA in the flower tubes, sesquiterpene emission from the flower tubes and accumulation of sesquiterpenes in pistils.

**RNA isolation and qRT-PCR analysis**
Total RNA was isolated from wild-type and *PhTPS1*-RNAi flower organs harvested at 14:00 h at different development stages using the Spectrum Plant Total RNA Kit (Sigma-Aldrich, St Louis, MO, USA). Total RNA samples were treated with DNaseI (Thermo Fisher Scientific, Waltham, MA, USA) and were reverse transcribed using the 5X All-In-One RT MasterMix (Applied Biological Materials Inc, Canada). Gene expression was analyzed by qRT-PCR using gene-specific primers (Supplementary Table 1) and the $E^{\Delta Ct}$ method [38]. Expression levels were normalized to the petunia ortholog of the known Arabidopsis reference gene *Protein Phosphatase 2A-like* [39,40].

**Metabolite profiling of internal pool and emission of terpenes**
For the analysis of internal pool of terpenes over development stages, flower organs from three to five flowers per biological replicate were harvested at 14:00 h and crushed in 3 mL hexane. Naphthalene (2 nmol) was added as internal standard (IS). Samples were vortexed 20 s, sonicated for 10 min and centrifuged at 2000 x *g* for 4 min. Supernatant was recovered, concentrated under nitrogen to approximately 200 µL and analyzed by GC-MS (see below).

For analysis of glycosides, tissues were collected at 14:00 h, crushed in MeOH and extracted overnight at -20°C. After sonication for 10 min, samples were centrifuged at 2500 x *g* for 5 min. Supernatant was recovered and split into two equal parts, before drying in Speedvac. Both samples were resuspended in 500 µL phosphate-citrate buffer (100 mM, pH 5) and 100 µL Viscozyme® L (Sigma-Aldrich, St Louis, MO, USA) was added to only one sample. Both samples were overlaid with 500 µL of hexane containing internal standard and incubated at 37°C overnight with a gentle shaking (120 rpm). The organic phase was then recovered, and extraction was repeated with additional 500 µL hexane. Hexane extracts from each sample were pooled together, concentrated under nitrogen to approximately 200 µL and analyzed by GC-MS (see below).

For the analysis of the effect of petunia tube on the terpene internal pools in pistils, tubes were removed from young flowers with scissors 4 days before anthesis. Naked pistils were covered with Eppendorf tubes to prevent desiccation. Sesquiterpene internal pools were analyzed in pistils after 4 days, which corresponds to day 1 of petunia flower development, as described above.

For the analysis of terpene content on pistil surface versus internal pools, pistils harvested on day 1 post-anthesis, were first dipped for 30 s in 1 mL hexane to extract cuticular terpenes, and then subjected to extraction of the remaining internal terpene pools as described above. Both samples were analyzed by GC-MS in parallel. Concurrent extraction of total terpenes from intact pistils was used as a control to ensure complete extraction during separate analysis of terpenoids on surface versus inside pistils. Each biological sample consisted of 5 pistils.

Petunia tube terpene emission was analyzed by dynamic headspace collection of volatiles starting at 12:00 h on day 0 before anthesis. Tubes with reproductive organs and petal limbs removed were placed in 5% sucrose solution in a sealed 1-liter glass jar with an inlet and an outlet. Cartridge containing 200 mg of Poropaq Q (80-100 mesh) (Sigma-Aldrich, St Louis, MO, USA) was placed in the inlet to purify incoming air. A volatile collection trap (VCT) filled with 50 mg of Poropak Q was inserted in the outlet. Volatiles were trapped on the VCT at a flow rate of 100 mL/min for 24 h. VCTs were eluted with 200 µL of dichloromethane containing 2 nmol naphthalene as an IS before GC-MS analysis.

For the analysis of terpene emission from the inner (adaxial) and outer (abaxial) surfaces of the petunia tube, tubes on day 0 before anthesis were harvested at 12:00 h, removed from their reproductive organs, fully opened to render the surface flat and placed in glass vials in a 5% sucrose solution. For Stir Bar Sorptive Extraction (SBSE), magnetic twister coated with Polydimethylsiloxane (PDMS) (Gerstel, Germany) were placed on the top of both sides of the tube and vials were closed. After 24 h, Twisters were eluted with 150 µL dichloromethane containing 2 nmol of IS and samples were analyzed by GC-MS. Emission of each terpene from the inner and outer side of the tube was presented as percentage of total emission of each terpene from both sides.

**Heterologous expression of terpene synthases in yeast**
To obtain full-length coding sequence (CDS) of petunia terpene synthases, total RNA was extracted from tubes (for *PhTPS1*) or pistils (for *PhTPS2*, *PhTPS3* and *PhTPS4*) of flowers harvested at 14:00 h on day 0 before anthesis using the Spectrum Plant Total RNA Kit (Sigma-Aldrich, St Louis, MO, USA). After treatment with DNase I (Thermo Fisher Scientific, Waltham, MA, USA), cDNAs were synthesized from 2 µg of total RNA with SuperScript III reverse transcriptase and oligo(dT)18 primers (Thermo Fisher Scientific, Waltham, MA, USA) according to the manufacturer's protocol. Vector constructs were generated as previously described [5] by using the USER cloning method (New England Biolabs) according to Nour-Eldin et al. [41]. USER extensions were added to the gene-specific primers (Supplementary Table 1) for subcloning the *PhTPS1*, *PhTPS2*, *PhTPS3* and *PhTPS4* coding sequences into the yeast expression plasmid pYeDP60u2.

To identify product specificity of sesquiterpene synthase candidates, the yeast expression constructs containing *PhTPS1*, *PhTPS3* and *PhTPS4* coding sequences were transformed into the WAT11 yeast strain. Yeast were cultured according to Pompon et al. [42] with few modifications as follows. After verification by PCR, individual transformed colonies were grown in selection medium overnight, then diluted 10-fold in 250 mL complete media and grown for another 30 h before induction with galactose (20 g/L). Induced cultures were poured in 1L glass bottle. An inlet tube was submerged in the culture to provide aeration, and a vacuum line with a VCT containing 200 mg of Poropak Q (80-100 mesh) was inserted in the outlet to trap produced terpenoids. The culture was stirred with a magnetic bar, and VOCs were collected for 2 days. Terpenoid products were eluted from the VCTs with 1 mL dichloromethane every 24 hours and analyzed by GC-MS. To determine product specificity of the monoterpene synthase candidate, the yeast expression construct containing *PhTPS2* coding sequence was transformed into the K197G yeast strain. Yeast were cultured according to Fischer et al. [43] with same modifications as mentioned above and VOC products were analyzed as above for the sesquiterpenes.

**GC-MS analysis**
GC-MS analysis was performed on an Agilent 7890B gas chromatograph (GC) (Agilent Technologies, Santa Clara, CA) equipped with a HP-5MS column (30 m, 0.25 mm, 0.25 μm; Agilent Technologies) and coupled to an Agilent 5977B high efficiency electro impact mass spectrometer (Agilent Technologies, Santa Clara, CA). 2 µL of sample were injected at 1:10 split using a Gerstel cooled injector system (CIS4, Gerstel, Germany) with an injection gradient of 12° per sec from 60°C to 250°C. Column temperature was held at 50°C for 0.5 min, then heated to 320°C (held for 5 min) at 20°C min$^{-1}$. Helium was used as a carrier gas at a flow rate of 1 mL min$^{-1}$. MS ionization energy was set at 70 eV, and the mass spectrum was scanned from 50 to 300 amu. For terpene profiling, chromatograms were analyzed with AMDIS software (http://www.amdis.net/) for mass spectra (MS) deconvolution using extracted ion chromatograms of specific ions 69 m/*z* (geraniol), 93 and 161 m/*z* for mono- and sesqui-terpenes and 128 m/*z* for IS, naphthalene. Products were identified by comparison of MS to the NIST/EPA/NIH Mass Spectral Library (Version 2.2). Quantification of terpenes was performed using the Mass Hunter quantitative software (Agilent Technologies) using response factors relative to the IS determined experimentally for the commercially available authentic standards germacrene D (representative sesquiterpene), geraniol (representative monoterpenol), and nerolidol (representative sesquiterpene alcohol) and normalized to the weight of the tissue.

**Terpene synthase activities in crude extracts from petunia tubes and pistils**
100 mg of the top 1 cm of tubes or whole pistils without ovaries were harvested at 14:00 h on day 0 before anthesis, frozen in liquid nitrogen, ground to a fine powder and resuspended in 1 ml of extraction buffer consisting of 50 mM 3-[N-morpholino]-2-hydroxypropanesulfonic acid, pH 6.9, 5 mM DTT, 5 mM $Na_2S_2O_5$, 1% [w/v] polyvinylpyrrolidone-40, and 10% glycerol, as previously described [44]. Extracts were shaken gently for 30 min on ice followed by centrifugation (15,000 x *g* for 20 min at 4°C). The supernatant was concentrated using Amicon® Ultra 0.5 mL Centrifugal Filters (Ultracel – 10K, MilleporeSigma, Darmstadt, Germany), desalted with Econo-Pac10DG Desalting Columns (Biorad) and resuspended in the assay buffer containing 50 mM HEPES, pH 7.2, 100 mM KCl, 7.5 mM $MgCl_2$, 20 µM $MnCl_2$, 5% (v/v) glycerol, 5 mM DTT. Protein concentration was determined by the Bradford method [45] with BSA as a standard. Activity assays were performed in a final volume of 500 µL assay buffer with 300 µg total protein and 100 µM (*E,E*)-FPP (*E*–*E*-farnesyl diphosphate) (Echelon Biosciences Inc, Salt-Lake City) as substrate. Assay mixture was overlayed with 500 µL hexane and incubated for 1 h at 28°C. Samples were extracted by adding 1mL of hexane containing 2 nmol of IS and vortexing. The organic phase was recovered, concentrated under nitrogen gas to about 200 µL and analyzed by GC-MS.

**Labeling experiments**
In order to track the tube-synthesized sesquiterpenes in pistils, wild-type flowers were harvested at day 0 before anthesis at 14:00 h, the tube was cut along the length and after removal of reproductive organs was placed in 1 mL 1% sucrose containing 5 mg of [2-$^{13}$C]-mevalonolactone in 20 mL scintillation vials (see experimental setup in Supplementary Fig. 10b). For each pistil, 1/3 of the style was removed and the pistil was placed in a microvial with 5% sucrose before returning back to the flower tube. The microvial ensured that pistils were not in contact with the labeled precursor. After 24 h, sesquiterpenes were extracted from both tubes and pistils and analyzed by GC-MS as described above. The amount of unlabeled and labelled sesquiterpenes were determined based on the specific ions 161 m/z (M + 0) and 162 + 163 m/z (M + 1 and M + 2), respectively.

**Exogenous fumigation**
In order to analyze the capacity of pistils to absorb exogenous sesquiterpenes, pistils were harvested on day 0 before anthesis at 14:00 h and placed in a small container with 1 mL 5% sucrose solution inside a 20 mL scintillation vials. A filter paper spotted with 10 µL of 1M caryophyllene (10 µmol) in methanol was placed in the scintillation vials, which were then capped. Three pistils were used for each biological replicate. Pistils were collected at the indicated time points after treatment and internal pools were extracted and analyzed as described above. To determine the relative ability of pistils, stamens and tubes to absorb sesquiterpenes, flowers were collected at day 0 before anthesis at 14:00 h, the tubes were opened with a blade along their length so that that each organ was exposed to the air, and placed in a small container with 1 mL 5% sucrose solution inside a 20 mL scintillation vials containing a filter paper spotted with 10 µL of 1M caryophyllene (10 µmol) as above. After 5 h of treatment, each organ was collected, extracted and analyzed for internal pools of caryophyllene as described above.

**Analysis of pistil parameters and seeds production**
Pistils from wild-type and *PhTPS1*-RNAi lines were harvested at 1 day post-anthesis at 14:00 h and weighed. The stigmas were dissected for imaging on a Leica MZ FLIII fluorescence stereo microscope (Leica Microsystems Inc., Buffalo Grove, IL). Major and minor axis lengths (see details in Supplementary Fig. 14d) were determined using ImageJ version 2.0.0-rc-68/1.52e. For measuring style diameter, hand sections were made just under the stigma, then imaged and analyzed as described above. To measure seed production, flowers from wild-type and *PhTPS1*-RNAi lines were emasculated on day 1 post-anthesis and cross-pollinated at 18:00 h on day 2 post-anthesis. 4 to 5 weeks after pollination, mature seed pods (brownish in color) were placed in 12 mL plastic tube. After pods cracked, seeds were collected and dried at 37°C for 2 days. Weight of total seeds per pod was determined, and 120 seeds per pod were counted out and weighed to determined seed weight and total number of seeds per pod.

**Microbiome profiling**
Pistils were removed from flowers using sterile forceps, taking care that pistils did not touch any other plant tissue. Pistils of wild type and *PhTPS1*-RNAi-11 plants were harvested on day 0 before anthesis and day 2 post-anthesis at 14:00 h and placed in ZR BashingBeads Lysis tubes containing 750μL of ZymoBIOMICS lysis solution (ZymoBIOMICS ™ DNA Miniprep Kit, Zymo Research Corporation, Irvine, CA, USA). Lysis tubes were shaken (20 Hz, 5 min) using TissueLyser II bead mill (Qiagen, Hilden, Germany) to extract DNA from microbes while leaving plant tissues undamaged. Microbial DNA was purified using the same kit following the manufacturer's instructions. Microbiome profiling of isolated DNA samples was performed by Eurofins Genomics (Ebersberg, Germany). Eurofins Genomics amplified and Illumina MiSeq sequenced the V3-V4 region of the 16S rRNA gene to identify bacterial OTUs (Operational Taxonomic Units and the ITS2 gene for fungal strains) following the standard procedure "InView - Microbiome Profiling 3.0 with MiSeq". Sequences were demultiplexed, the primers were clipped, forward and reverse reads were merged, and merged reads were quality filtered. Microbiome analysis was performed by Eurofins Genomics using the company's standard procedure (the following description of analysis is provided by Eurofins Genomics): reads with ambiguous bases ("N") were removed. Chimeric reads were identified and removed based on the de-novo algorithm of UCHIME [46] as implemented in the VSEARCH package [47]. The remaining set of high-quality reads was processed using minimum entropy decomposition [48,49]. Minimum Entropy Decomposition (MED) provides a computationally efficient means to partition marker gene datasets into OTUs (Operational Taxonomic Units). Each OTU represents a distinct cluster with significant sequence divergence to any other cluster. By employing Shannon entropy, MED uses only the information-rich nucleotide positions across reads and iteratively partitions large datasets while omitting stochastic variation. The MED procedure outperforms classical, identity-based clustering algorithms. Sequences can be partitioned based on relevant single nucleotide differences without being susceptible to random sequencing errors. This allows a decomposition of sequence data sets with a single nucleotide resolution. Furthermore, the MED procedure identifies and filters random "noise" in the dataset, i.e. sequences with a very low abundance (less than 0.02% of the average sample size). To assign taxonomic information to each OTU, DC-MEGABLAST alignments of cluster representative sequences to the sequence database were performed (Reference database: NCBI_nt (Release 2018-07-07)). A most specific taxonomic assignment for each OTU was then transferred from the set of best-matching reference sequences (lowest common taxonomic unit of all best hits). Hereby, a sequence identity of 70% across at least 80% of the representative sequence was a minimal requirement for considering reference sequences. Further processing of OTUs and taxonomic assignments was performed using the QIIME software package (version 1.9.1, http://qiime.org/) [50]. Abundances of bacterial taxonomic units were normalized using lineage-specific copy numbers of the relevant marker genes to improve estimates [51].

Given the low diversity of bacterial OTUs in the samples and the high number of sequences that were assigned to be originating from plant tissues, we compared the microbiome of the pistils to the floral microbiome of a *Brassica rapa* plant that was cultivated from surface sterilized seeds under sterile conditions and of *Brassica rapa* plant cultivated in soil in the lab (Dataset S1). For detailed methods see [23]. Whereas the microbiome of the flowers of plants cultivated from surface-sterilized seeds consisted of three OTUs, the microbiome of the two samples of non-sterile flowers was more diverse with 68 and 115 OTUs. As most samples contained an OTU belonging to the genus Delftia, which is likely a contaminant, these OTUs were removed from analysis. The other OTUs found to be associated with petunia pistils were not found in sterile *Brassica* samples. The comparison of the microbial communities of the WT and *PhTPS1*-RNAi-11 petunia flowers with the microbial communities of *B. rapa* indicates that the relatively low number of reads assigned to bacterial OTUs in the petunia samples reflects the low diversity and abundance of bacteria on these surfaces rather than a methodological artifact. However, further experiments may be required to fully characterize the microbiome of petunia pistils. Fastaq files of samples containing the sequences of the OTUs associated with *Petunia* and *Brassica* are deposited at the European Nucleotide Archive (PRJEB29416 (ERP111715)).

**Analysis of pollen tube growth**
Flowers from wild-type and the *PhTPS1*-RNAi-11 line were emasculated day 1 post-anthesis and cross-pollinated at 18:00 on day 2 post-anthesis. The pollinated pistils were then collected at 5 h, 15 h, 25 h and 36 h after hand pollination and fixed immediately in ethanol-acetic acid (3:1 v/v) for 2 h. After clearing in 8 N NaOH at room temperature for 24 h, they were stained using aniline blue (0.1% aniline blue in $K_3PO_4$) for 24 h [52]. Four pistils per sample were analyzed for each time point and images were captured with a Nikon Eclipse Ti2-E microscope (Nikon Instruments Inc, Melville, NY).

**Complementation assays**
For gas phase complementation of terpenes in *PhTPS1*-RNAi pistils, flowers from wild-type and *PhTPS1*-RNAi-11 line were harvested at day 0 before anthesis at 12:00 h, the tube was cut along the length and after removal of reproductive organs was placed in 1 mL 1% sucrose in 20 mL scintillation vials (see experimental setup in Supplementary Fig. 10c). For each pistil, 1/3 of the style was removed and the pistil was placed in a microvial with 5% sucrose before returning back to the flower tube. Wild type pistils were placed in wild type tubes, and *PhTPS1*-RNAi-11 pistils were placed either in wild type tubes or *PhTPS1*-RNAi-11 tubes. After 24 h, sesquiterpenes were extracted from pistils and analyzed by GC-MS as described above.

In order to complement the development of *PhTPS1*-RNAi-11 stigma, pistils from wild-type and *PhTPS1*-RNAi-11 line were harvested 4 days before anthesis at 12:00 h, and grown *in vitro* in Magenta boxes containing Murashige and Skoog (MS) medium basal salt (Sigma-Aldrich, St Louis, MO, USA) supplemented with 3% sucrose, 0.01 mg/L thiadiazuron and 0.1 mg/L gibberellic acid (see experimental setup in Supplementary Fig. 10d). In parallel, wild-type and *PhTPS1*-RNAi-11 line flowers were collected on day 0 before anthesis at 12:00 h, surface sterilized for 2 min with 50% commercial bleach and rinsed several times with sterile water. Tubes were cut along their length and after removal of reproductive organs were carefully placed around the pistils in MS media. Wild type tubes were either placed around wild type or *PhTPS1*-RNAi-11 pistils, and *PhTPS1*-RNAi-11 tubes were either placed around wild type or *PhTPS1*-RNAi-11 pistils. Tubes were replaced every 48 h. *In vitro* reassembled flowers were cultivated at 20ºC under fluorescent light (35 µE $m^{-2}s^{-1}$) with a 16 h/8 h light/dark photoperiod. After 4 days, pistils were harvested, weighed and the stigma and style parameters (see details in Supplementary Fig. 14d) were measured using a microscope as described above.

**Quantification of total wax from tube and pistil**
Wild-type tubes, pistils, and stamen were harvested on day 0 before anthesis at 14:00 h and separately submerged in glass scintillation vials each containing 5 mL of hexane. Vials were vortexed for 30 sec and the solvent was decanted into clean scintillation vials. Hexane extracts were subsequently dried to completion under a gentle stream of nitrogen gas. Dried wax residues were then weighed to quantify the total wax coverage on individual floral components. Each biological replicate consisted of floral organs harvested from twelve day 0 flower buds.

**Data availability**

The data that support the findings of this study are available from the corresponding author upon reasonable request. Plant material generated in this study is available from the corresponding author upon request. For the microbiome: Fastaq files of samples containing the sequences of the OTUs associated with *Petunia* and *Brassica* are deposited at the European Nucleotide Archive (PRJEB29416 (ERP111715)). The sequences reported in this paper have been deposited in GenBank database with the following accession numbers: MK159027 for PhTPS1, MK159028 for PhTPS2, MK159029 for PhTPS3, and MK159030 for PhTPS4.

**Methods-only References**

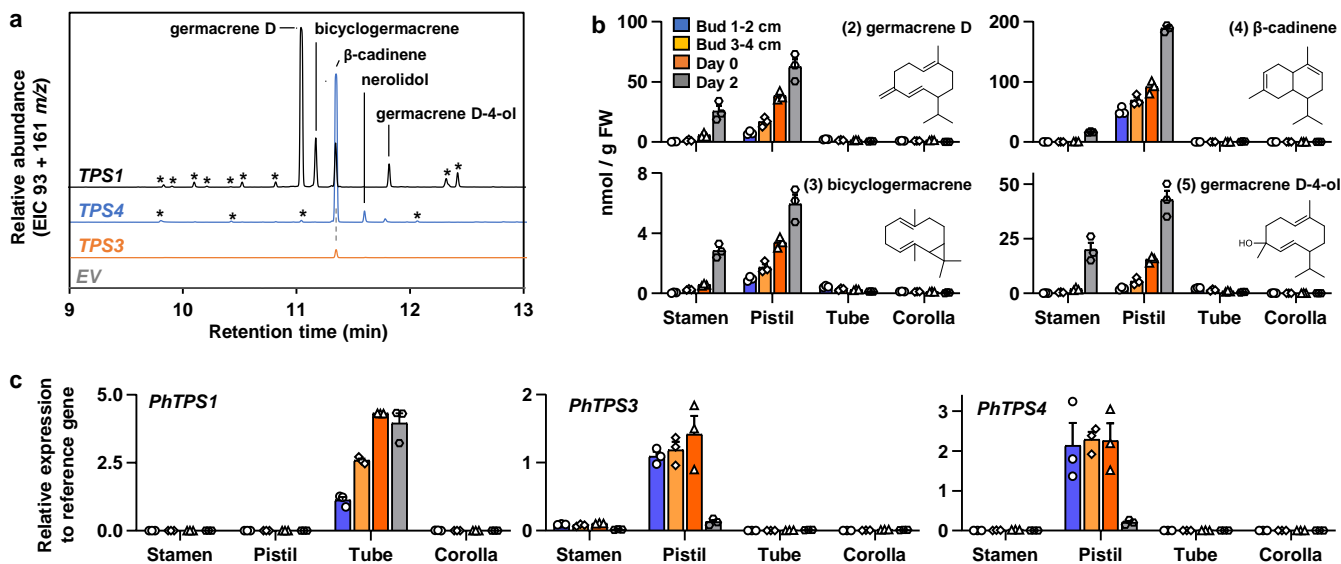

**Fig. 1 | Characterization of sesquiterpene synthases expressed in petunia flowers.**
(**a**) Products of PhTPS1, PhTPS3 and PhTPS4 or empty vector (EV) expressed in yeast strain WAT11. GC-MS chromatograms of volatiles emitted from yeast cultures are presented as total extracted ion current (EIC, m/z 93 + 161). Sesquiterpenes were identified by comparison of their mass spectra (MS) to the NIST library. Germacrene D was confirmed by comparison with an authentic standard. Asterisks represent unidentified putative sesquiterpenes. GC-MS chromatograms are representative of three independent experiments. (**b**, **c**) Internal pools of sesquiterpenes (**b**) and relative transcript levels of *PhTPS1*, *PhTPS3* and *PhTPS4* (**c**) in petunia stamen, pistil, tube and corolla over flower development (1-2 cm buds, 3-4 cm buds, flower day 0 before anthesis, and flower day 2 after anthesis) determined by GC-MS (**b**) and by qRT-PCR (**c**), respectively. Internal pools were quantified based on the ratio of the integrated peak area of terpenoids relative to the IS peak area and normalized to the weight of the tissue. *PhPP2A* was used as a reference gene in qRT-PCR. Data are means ± SE (n = 3 biological replicates).

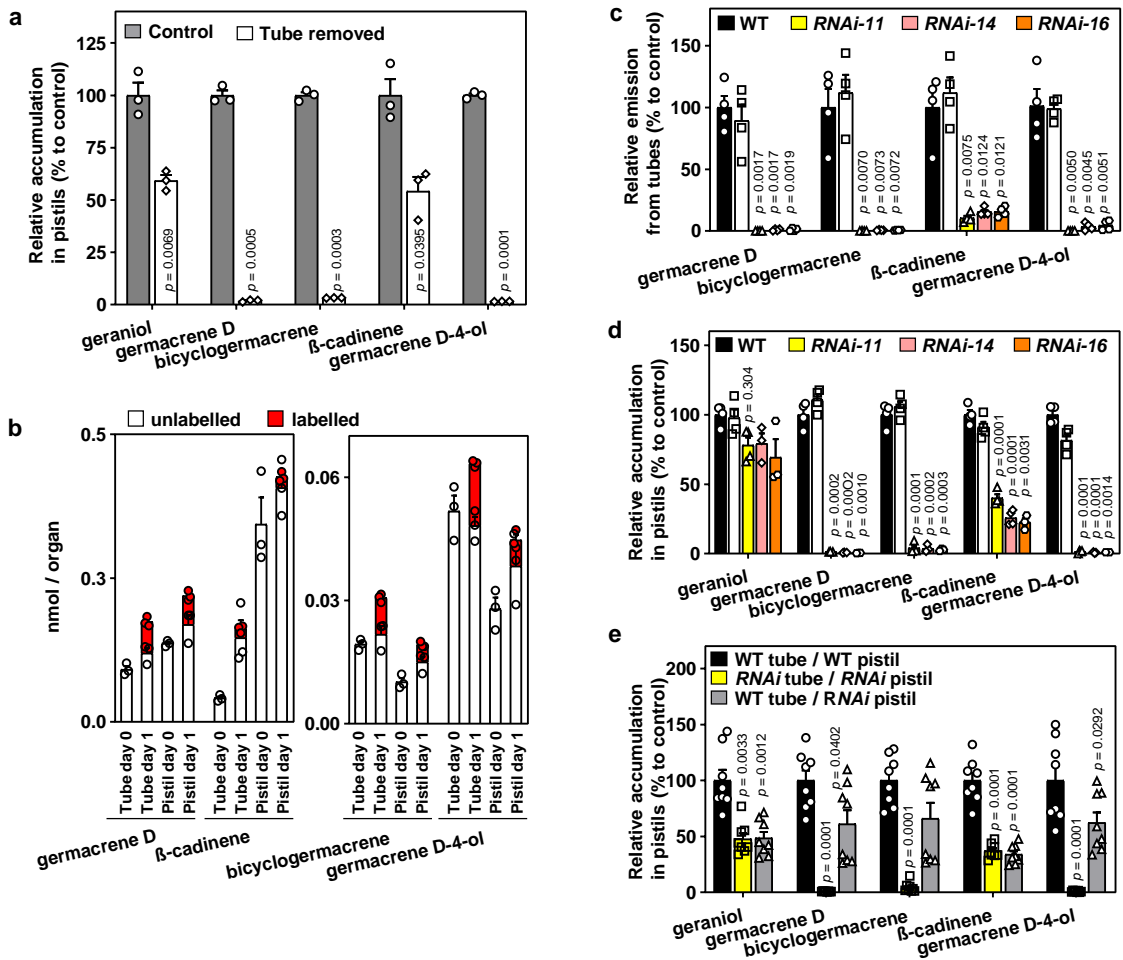

**Fig. 2 | Analysis of inter-organ transport of PhTPS1 products in petunia buds.**

(**a**) Effect of flower tube on accumulation of terpenoids in pistils. Terpenoids were analyzed in pistils on day 1 post-anthesis from intact flowers (control) and from flowers from which tubes were removed 4 days before anthesis. Data are means ± SE (n = 3 biological replicates). (**b**) GC-MS analysis of sesquiterpenes produced by petunia tubes fed with [2-$^{13}$C]-mevalonolactone for 24 h. Amounts of unlabeled and labelled sesquiterpenes were quantified with the specific ions 161 m/z and 162 + 163 m/z (M + 1 and M + 2), respectively. Data are means ± SE (n = 3 biological replicates). (**c** and **d**) Effect of *PhTPS1*-RNAi downregulation on terpenoid emission from tubes over 24 h beginning on day 0 before anthesis (**c**) and their accumulation in pistils on day 1 post-anthesis (**d**). WT, wild type control; EV, empty vector control, and *PhTPS1*-RNAi lines. Data are means ± SE (n = 4 biological replicates). (**e**) Effect of gas phase complementation by wild type tubes on internal pools of mono- and sesquiterpenes in *PhTPS1*-RNAi-11 pistils. Terpenoids were extracted from pistils after 24 h of complementation and analyzed by GC-MS. Data are means ± SE (n = 8 biological replicates). Experimental setups for (**a**), (**b**) and (**e**) are illustrated on Supplementary Fig. 10. Significant p values (p < 0.05), as determined by two-tailed paired t-test relative to control, are shown.

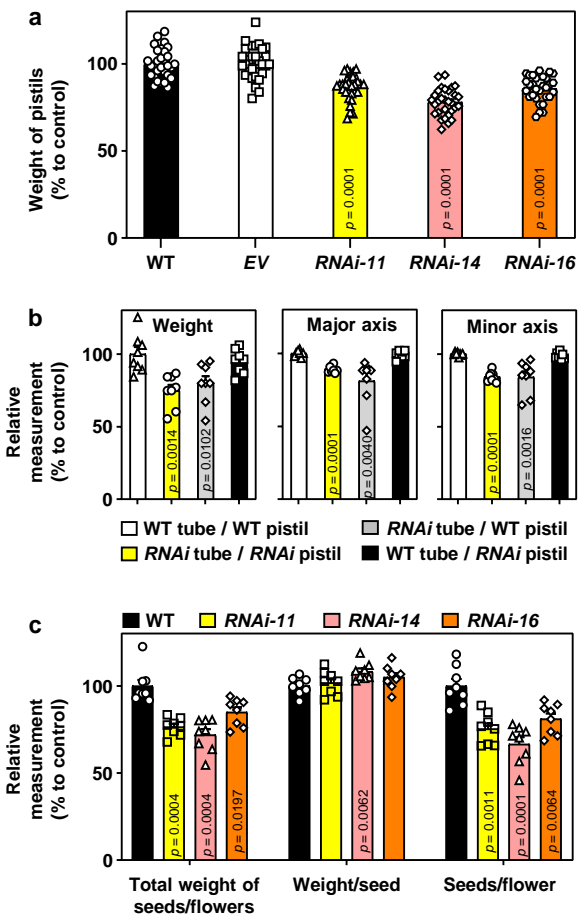

**Fig. 3 | Effect of sesquiterpene fumigation on pistil development and seed yield.**

(**a**) Weight of pistils from wild type (WT), an empty vector (EV) control line and *PhTPS1*-RNAi lines on day 1 post-anthesis. Data are means ± SE (n = 30 biological replicates). (**b**) Complementation via gas phase of in vitro growth of *PhTPS1*-RNAi-11 pistils with WT tubes. Experimental setup is illustrated on Supplementary Fig. 10d and described in Materials and Methods. After 4 days of complementation, weight and major and minor axes of stigma were measured. Data are means ± SE (n = 8 biological replicates). (c) Effect of *PhTPS1*-RNAi downregulation on seed production. Total weight of seeds per flower, weight per seed and number of seeds per flower were measured in WT and *PhTPS1*-RNAi transgenic lines. Data are means ± SE (n = 8 biological replicates). Significant p values (p < 0.05), as determined by two-tailed paired t-test relative to control, are shown.

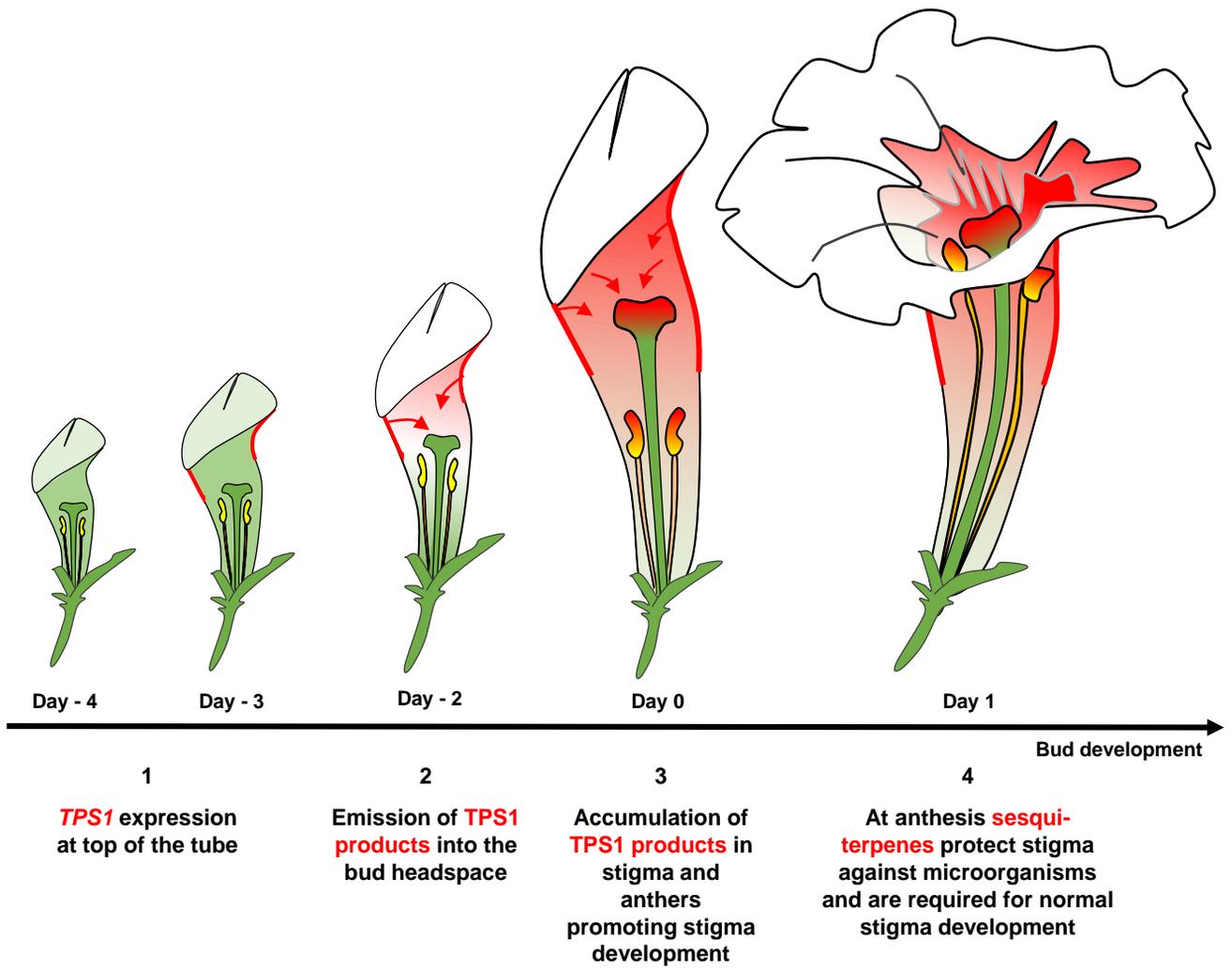

**Fig. 4 | Proposed fumigation model.**
Scheme showing the sesquiterpene fumigation process during petunia bud development. At young bud stage, *PhTPS1* is expressed at the top of the tube below the corolla (shown in red). PhTPS1 sesquiterpene products are emitted in the bud headspace (red area and arrows) and are then absorbed and accumulated by the stigma and anthers. This process ensures normal pistil development and protects reproductive organs against microorganisms when flower opens.

# Supplementary Materials for "Natural fumigation as a mechanism for volatile transport between flower organs"

Benoît Boachon, Joseph H. Lynch, Shaunak Ray, Jing Yuan, Kristian Mark P. Caldo, Robert R. Junker, Sharon A. Kessler, John A. Morgan & Natalia Dudareva.
correspondence to: dudareva@purdue.edu

**This PDF file includes:**
**Supplementary Figures 1 to 16,**
**Supplementary Table 1,**
**Captions for Dataset 1.**

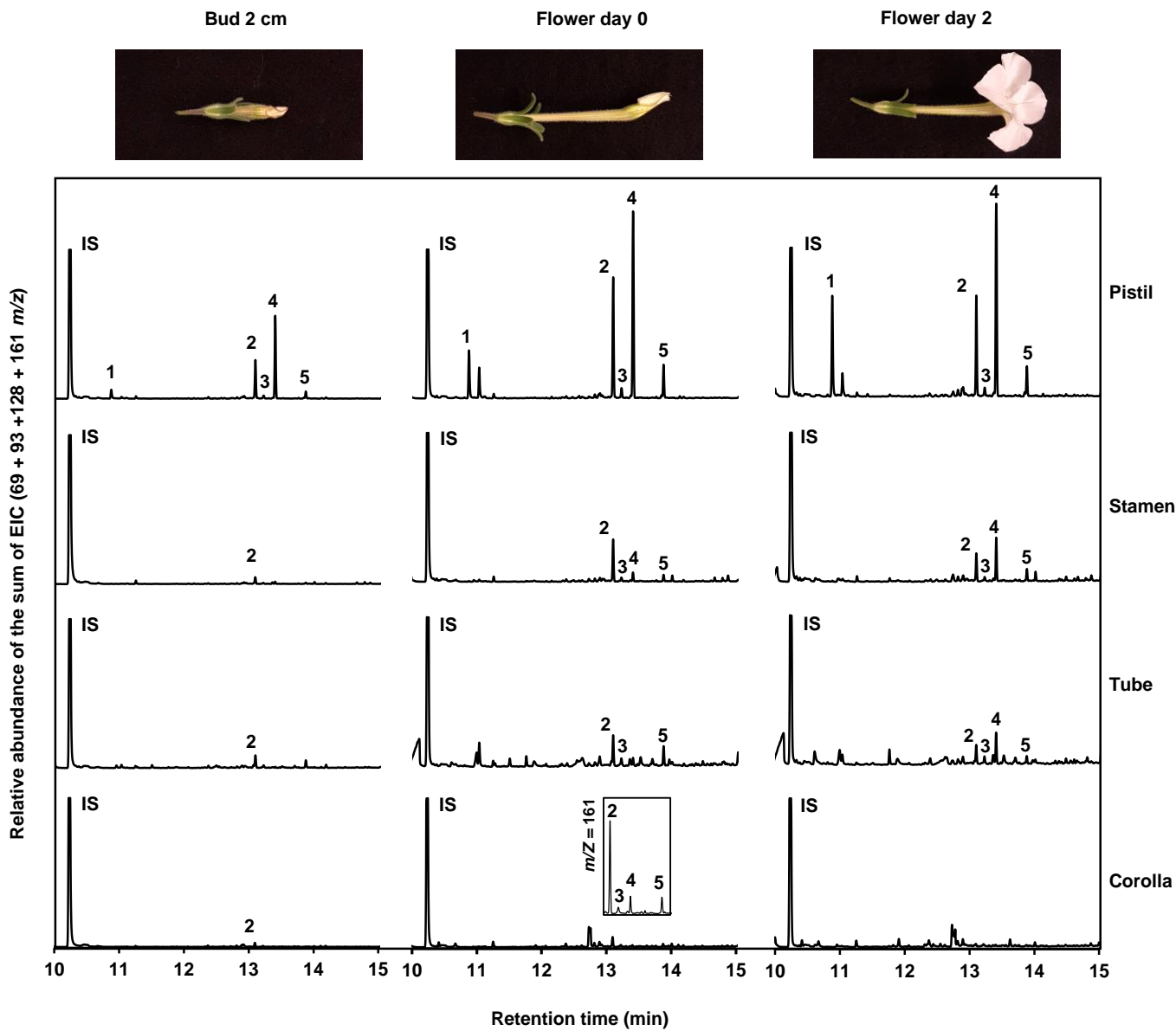

**Supplementary Fig. 1 | Metabolic profiling of internal terpene pools in petunia flower organs during development.**

GC-MS chromatograms of hexane extracts from petunia corolla, tube, stamen and pistil at different stages of flower development (2 cm buds, flower day 0 before anthesis, and flower day 2 after anthesis) showing total extracted ion current (EIC = $m/z$ 69 + 93 + 128 + 161). Inset in corolla flower day 0 shows EIC of $m/z$ = 161. Each sample contained organs collected from 5 flowers. Naphthalene was used as internal standard (IS). Mono- and sesquiterpenes were identified using NIST libraries and are numbered as follows: (1) geraniol, (2) germacrene D, (3) bicyclogermacrene, (4) β-cadinene, (5) germacrene D-4-ol. GC-MS chromatograms are representative of three independent experiments.

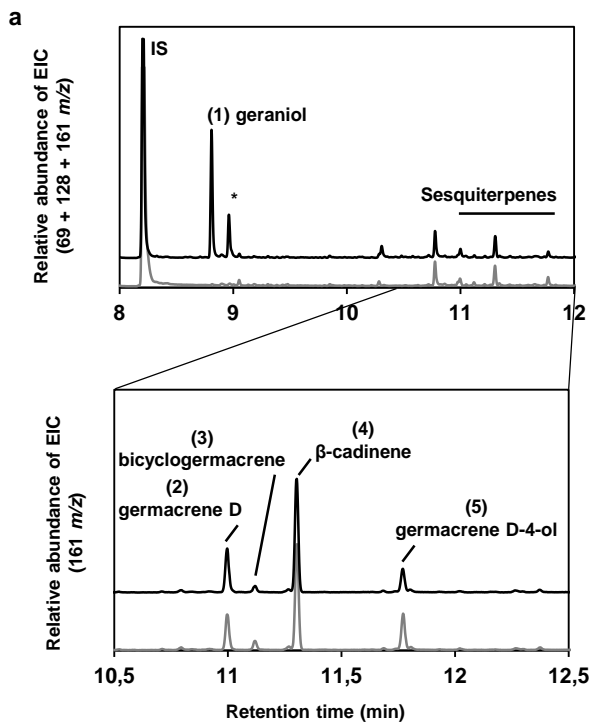

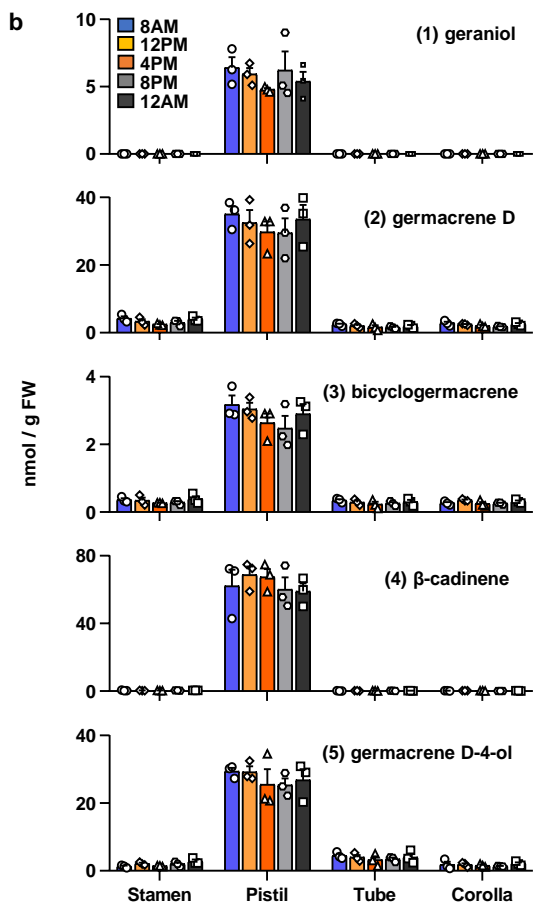

**Supplementary Fig. 2 | Analysis of terpenoids and their glycosides in petunia flowers.**
(**a**) GC-MS chromatograms showing the relative abundance of extracted ion current (EIC). Petunia pistils on day 1 after anthesis were extracted with MeOH, dried and treated (black) or not (grey) with Viscozyme to release aglycones from glycosides. Naphthalene was used as IS, geraniol and sesquiterpenes are shown on chromatograms and asterisk corresponds to an alcohol green leaf volatiles. GC-MS chromatograms are representative of three independent experiments. (**b**) Terpene internal pools in different floral tissues throughout the day as determined by GC-MS. Data are means ± SE (n = 3 biological replicates).

**a**

| Terpene synthaze ID Contig_ID | Average FPKM in bud ± SD (Average CPM ± SD) | Average FPKM in day 2 flowers ± SD (Average CPM ± SD) | Fold change (edgeR FC) | FDR |
|---|---|---|---|---|
| *PhTPS1* *comp2166* | 6,083.0 ± 1,138.7 (56.1 ± 10.0) | 3,532.3 ± 747.0 (38.9 ± 5.3) | 0.6 0.7 | 2.3E-02 |
| *PhTPS2* *comp2166* | 429.7 ± 31.5 (4.0 ± 0.5) | 249.0 ± 156.0 (2.7 ± 1.5) | 0.6 0.7 | 2.2E-02 |
| *PhTPS3* *comp19516* | 409.0 ± 47.8 3.8 ± 0.9 | 398.3 ± 58.9 4.4 ± 0.8 | 1.0 1.2 | 4.5E-01 |
| *PhTPS4* *comp254* | 7.0 ± 4.0 (0.06 ± 0.03) | 8.7 ± 4.2 (0.09 ± 0.04) | 1.2 1.5 | 4.3E-01 |

**b**

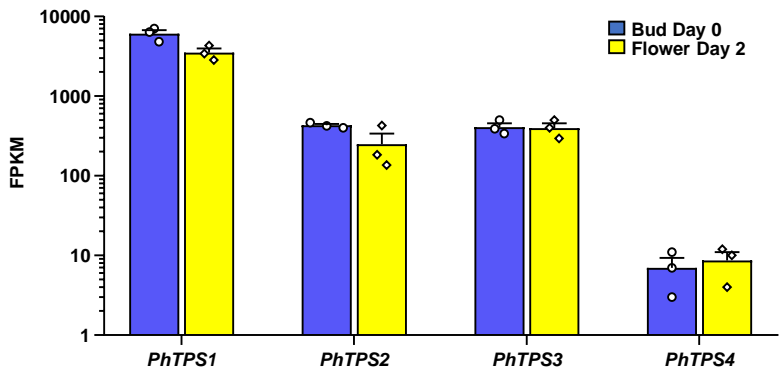

**Supplementary Fig. 3 | Terpene synthase candidates identified in petunia RNAseq datasets.**
(**a**) Table and (**b**) graphical overview of the transcript levels for four terpene synthase (*TPS*) candidates found to be expressed at the bud stage (day 0) and in flower day 2 post-anthesis. Gene expression was measured based on Cufflinks and edgeR analysis of RNAseq data[12]. Data represent average of FPKM (fragments per kilobase per million mapped fragments) and CPM (counts per million reads) from three biological replicates ± SD. Table additionally contains average fold changes and FDR (false discovery rate) values from edgeR analysis[12]. FDR was calculated using EdgeR by first applying an exact test to determine the *p*-value of differential expression, and then adjusting *p*-value using the Benjamini-Hochberg method to control for multi-testing error.

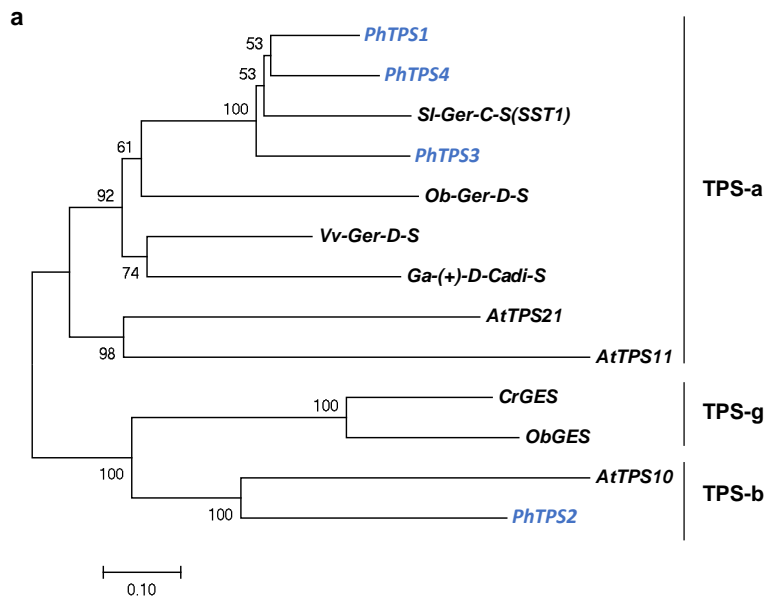

**Supplementary Fig. 4 | Phylogenetic analysis of terpene synthases expressed in petunia flowers and their matrix identity.**
(**a**) Neighbor-joining phylogenetic tree, including the four *Petunia hybrida* (Ph) terpene synthases identified in flowers by RNAseq (in blue), drawn using the MEGA 7 software (*42*). Tree also contains nine characterized terpene synthases from other species: the multi-sesquiterpene synthase AtTPS11 (Q4KSH9), the caryophyllene synthase AtTPS21 (Q84UU4) and the linalool synthase AtTPS10 (Q9ZUH4) from *Arabidopsis thaliana*; geraniol synthase (CrGES) from *Catharanthus roseus* (J9PZR5), ObGES from *Ocimum basilicum* (Q6USK1), germacrene D synthase (Ob-Ger-D-S) from *Ocimum basilicum* (Q5SBP6), germacrene C synthase (Sl-Ger-C-S) from *Solanum lycopersicum* (O64961), VvGer-D-S from *Vitis vinifera* (Q6Q3H3) and δ-cadinene synthase (Ga-δ-Cadi-S) from *Gossypium arboretum* (Q39761). The optimal tree with the sum of branch length equal to 4.76878608 is shown. Bootstrap percentages are shown next to the branches (1000 iterations). The tree is drawn to scale, with branch lengths in the same units as those of the evolutionary distances used to infer the phylogenetic tree. The evolutionary distances were computed using the Poisson correction method and are in the units of the number of amino acid substitutions per site. All positions containing gaps and missing data were eliminated. There was a total of 498 positions in the final dataset. TPS-a ,TPS-b and TPS-g clades are shown. (**b**) The percent identity matrix of the 13 TPSs computed using Clustal Omega (https://www.ebi.ac.uk/Tools/msa/clustalo/).

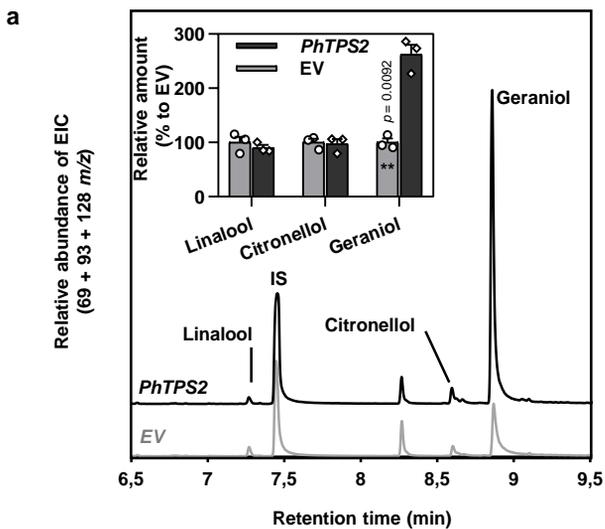

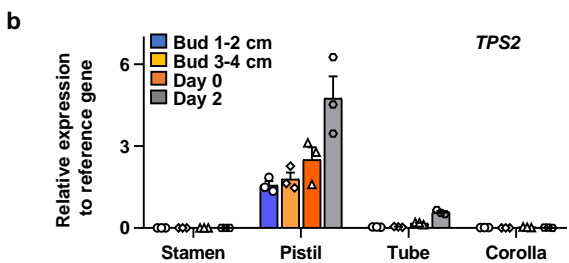

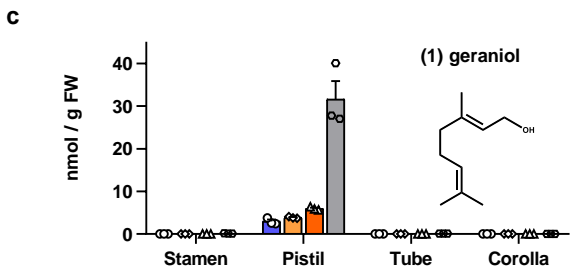

**Supplementary Fig. 5 | Characterization of PhTPS2.**
(**a**) Products of PhTPS2 or empty vector (EV) expressed in yeast strain K197G (*9*). GC-MS chromatograms of volatiles emitted from yeast cultures are presented as total extracted ion current (EIC, m/z 69 + 93 + 128). Naphthalene was used as internal standard (IS). As K197G strain produces linalool, citronellol and geraniol, inset shows the quantification of these monoterpenols. Data are means ± SE (n=3 biological replicates). Significant *p* value ($p < 0.01$), as determined by two-tailed paired *t*-test relative to EV is shown. (**b**) Transcript levels of *PhTPS2* in petunia stamen, pistil, tube and corolla tissues over flower development (1-2 cm buds, 3-4 cm buds, flower day 0 before anthesis, and flower day 2 after anthesis) determined by qRT-PCR relative to *PP2A* reference gene. Data are means ± SE (n = 3 biological replicates). (**c**) Geraniol internal pools in the same tissues as shown in (**c**) determined by GC-MS. Quantification was performed based on the ratio of the integrated peak area of geraniol relative to the IS peak area and normalized to the weight of the tissue. Data are means ± SE (n = 3 biological replicates).

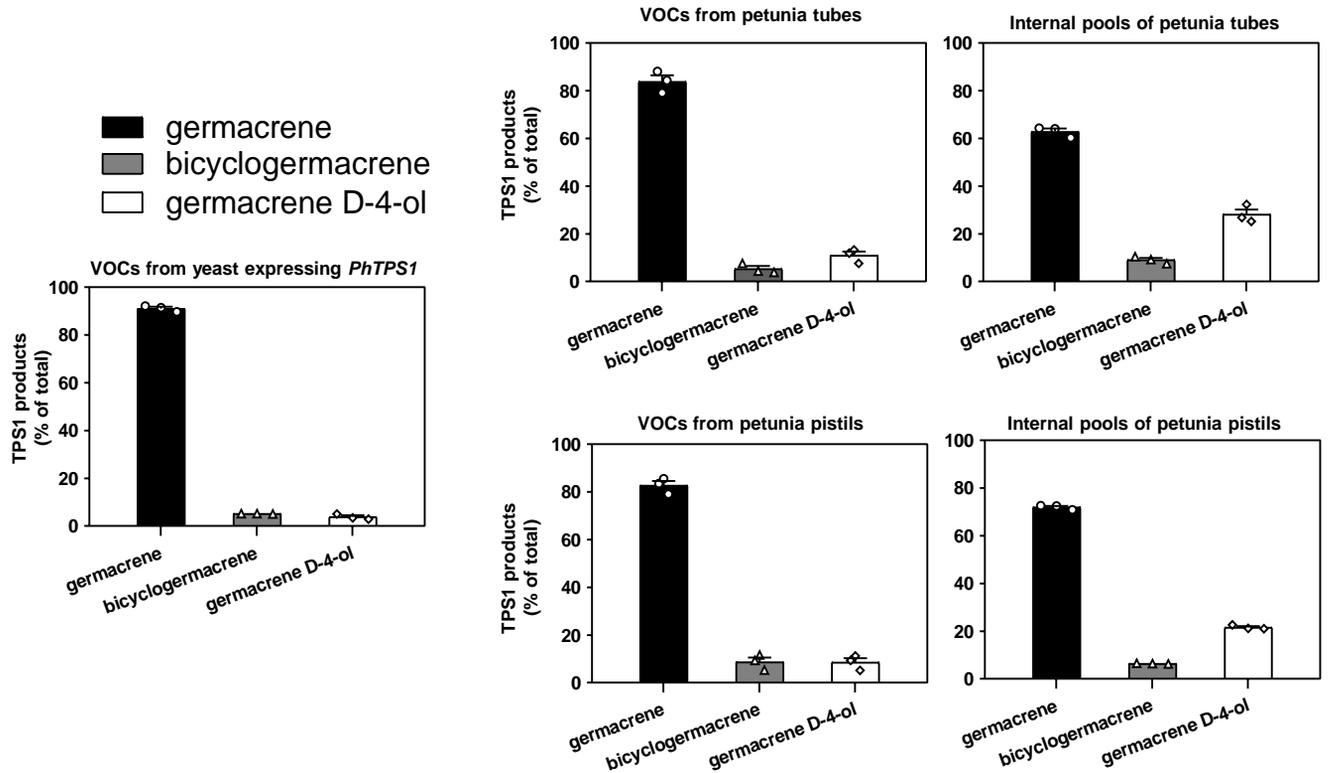

**Supplementary Fig. 6 | PhTPS1 products synthesized in plant tissues and by recombinant PhTPS1 protein.**
*PhTPS1* products germacrene, bicyclogermacrene and germacrene D-4-ol were quantified from headspace of yeast expressing *PhTPS1*, headspace from WT petunia tubes and pistils, and internal pools of WT petunia tubes and pistils. Data are shown as the percentage of total PhTPS1 products. Cadinene was omitted from this analysis since it is also the product of PhTPS3 and PhTPS4. Data are means ± SE of 3 biological replicates.

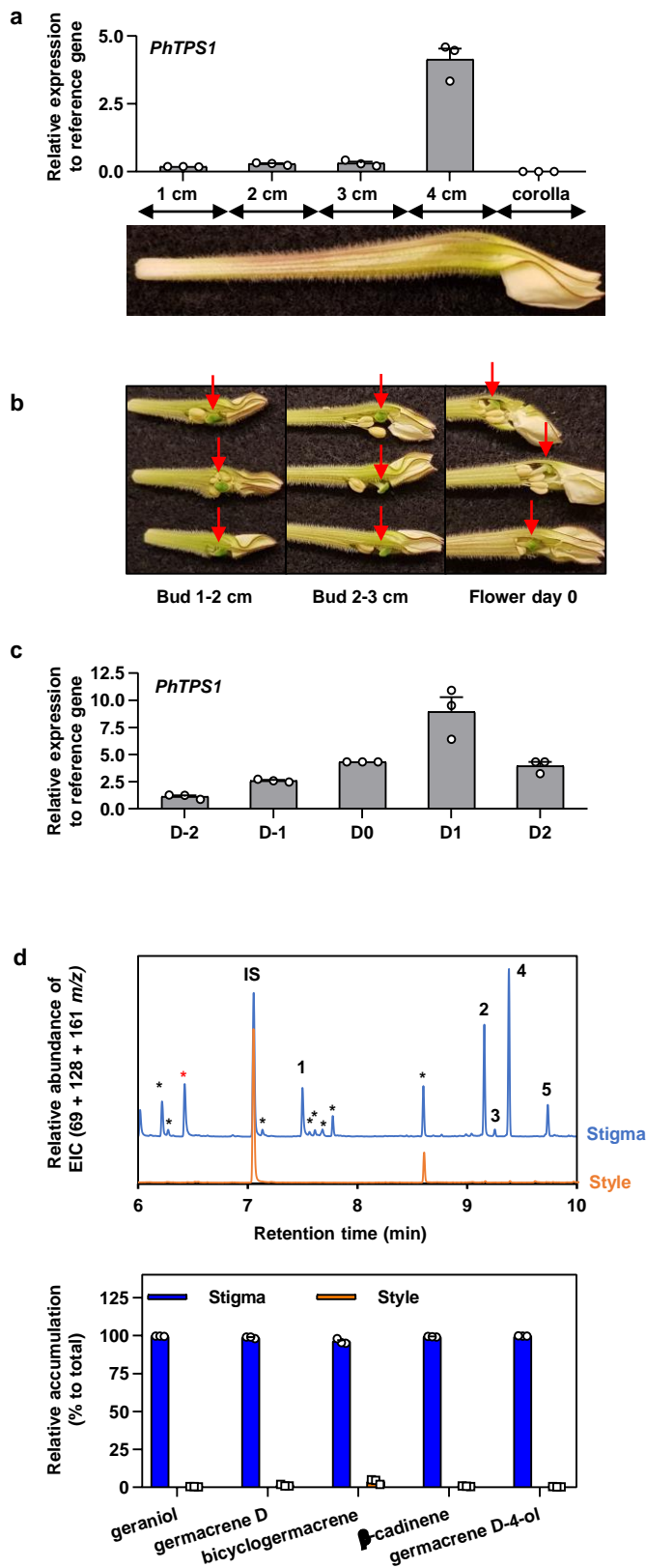

**Supplementary Fig. 7 | *PhTPS1* expression in petunia tube.**
(**a**) *PhTPS1* transcript levels in different parts of the tube on day 0 before anthesis determined by qRT-PCR relative to reference *PP2A* gene. (**b**) Pictures of dissected petunia buds at different developmental stages showing position of stigma in close proximity to the part of the tube where *PhTPS1* is expressed the most. Three individual flowers are shown for each developmental stage. (**c**) Relative *PhTPS1* transcript levels in the top part of the tube over flower development (D-2, D-1, D0, D1 and D2 are day -2, day -1 and day 0 before anthesis, and day 1 and day 2 after anthesis, respectively) determined by qRT-PCR relative to reference *PP2A* gene. (**d**) Distribution of accumulated terpenoids within the pistil. Representative GC-MS chromatogram of hexane extract of petunia stigma and style on day 0 before anthesis (upper panel) and quantification of the different terpenes in stigma and style (lower panel). Total amount in stigma and style set as 100%. Chromatograms show the relative abundance of the sum of extracted ion chromatograms (EIC = m/z 69 + 128 + 161). Naphthalene was used as internal standard. Mono- and sesquiterpenes present in samples were identified using NIST libraries and are numbered as follows: (1) geraniol, (2) germacrene D, (3) bicyclogermacrene, (4) β-cadinene, (5) germacrene D-4-ol, black asterisks represents different green leaf volatiles, red asterisk represent phenylethanol. All presented data are means ± SE (n = 3 biological replicates).

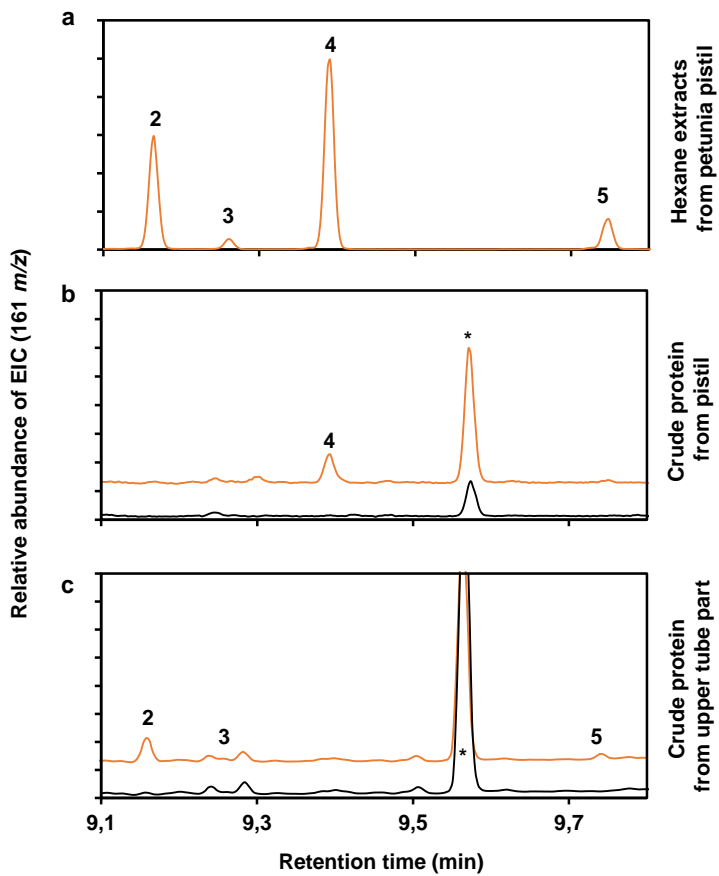

**Supplementary Fig. 8 | Products formed from farnesyl diphosphate by crude extracts prepared from petunia pistil and tube.**

(**a**) Representative GC-MS chromatogram showing the main sesquiterpenes present in hexane extracts of petunia pistil on day 0 before anthesis. Total crude proteins were extracted from pistils (**b**), and upper part of tubes (**c**) of petunia flowers on day 0 before anthesis. Assays were performed with 300 µg of either total proteins (orange) or boiled proteins as negative controls (black). Chromatograms in **b**, and **c** show the relative abundance of the EIC 161 *m/z* with normalized scale to each other. Products are numbered as follows: (2) germacrene D, (3) bicyclogermacrene, (4) β-cadinene, (5) germacrene D-4-ol, asterisk shows the presence of nerolidol as a degradation product of FPP. GC-MS chromatograms in **a** – **c** are representative of three independent experiments.

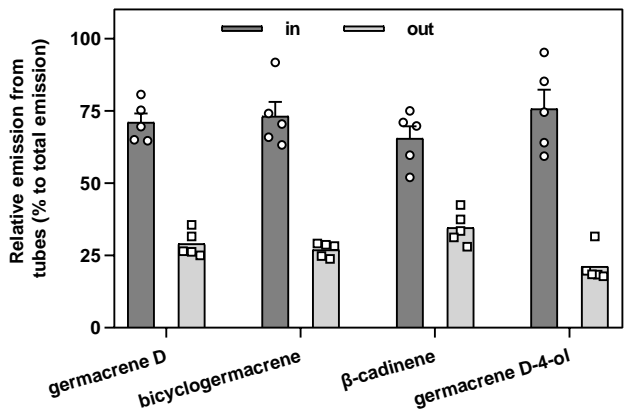

**Supplementary Fig. 9 | Emission of PhTPS1 products from the inner (adaxial) and outer (abaxial) surfaces of the petunia tube.**
VOCs were collected for 24 h using Twisters placed on the inner and outer sides of petunia tubes on day 0 before anthesis, and analyzed by GC-MS. Results are means ± SE (n = 5 biological replicates).

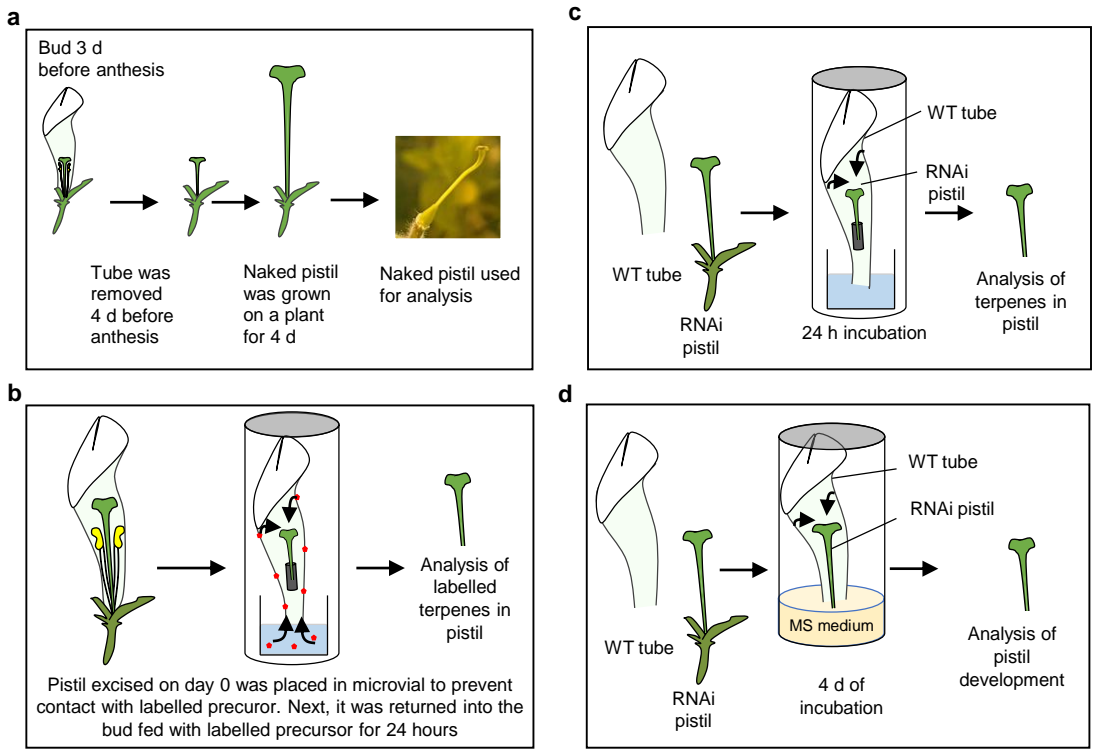

**Supplementary Fig. 10 | Experimental setups to test the aerial transport of TPS1 products from tube to pistil and its effect on pistil development by complementation.**
(**a**) Scheme illustrating the experimental setup to generate the results shown in Fig. 2a to test the effect of flower tube on the accumulation of terpenoids in pistils. Terpenoids were analyzed in pistils on day 1 post-anthesis from intact flowers (control) and from flowers that had tubes removed 4 days before anthesis. (**b**) Scheme illustrating the the experimental setup to generate the results shown in Fig. 2b to test the aerial transport of labelled sesquiterpenes between tubes and pistils. Petunia buds were harvested on day 0 before anthesis. Pistils were excised and placed in microvials to prevent their contact with the feeding solution containing the labelled precursor and then returned into the buds. Reconstituted buds were placed in glass vials with tubes inserted in a solution containing 2% sucrose and 1mg/mL [2-$^{13}$C]-mevalonolactone. Feeding was done for 24 h before analysis of terpenoid labeling in tubes and pistils. (**c**) Scheme illustrating the experimental setup to generate the results shown in Fig. 2e to test whether natural fumigation from wild type tubes could complement the lack of sesquiterpenes in *PhTPS1*-RNAi-11 pistils. WT and *PhTPS1*-RNAi-11 buds were harvested on day 0 before anthesis. Experiment was performed as in (**b**) but without labelled precursor and with different combinations of WT and *PhTPS1*-RNAi-11 tubes and pistils. (**d**) Scheme illustrating the experimental setup to generate the results shown in Fig. 3b to test whether sesquiterpene fumigation from wild type tubes could complement the development of *PhTPS1*-RNAi-11 pistils. WT and *PhTPS1*-RNAi-11 pistils were harvested 3 days before anthesis. WT and *PhTPS1*-RNAi-11 tubes were harvested on day 0 before anthesis and removed from their reproductive organs. Reconstituted buds, consisting of petunia tubes and pistils in different genotype combinations, were grown in MS media for 4 days before analysis of pistil development.

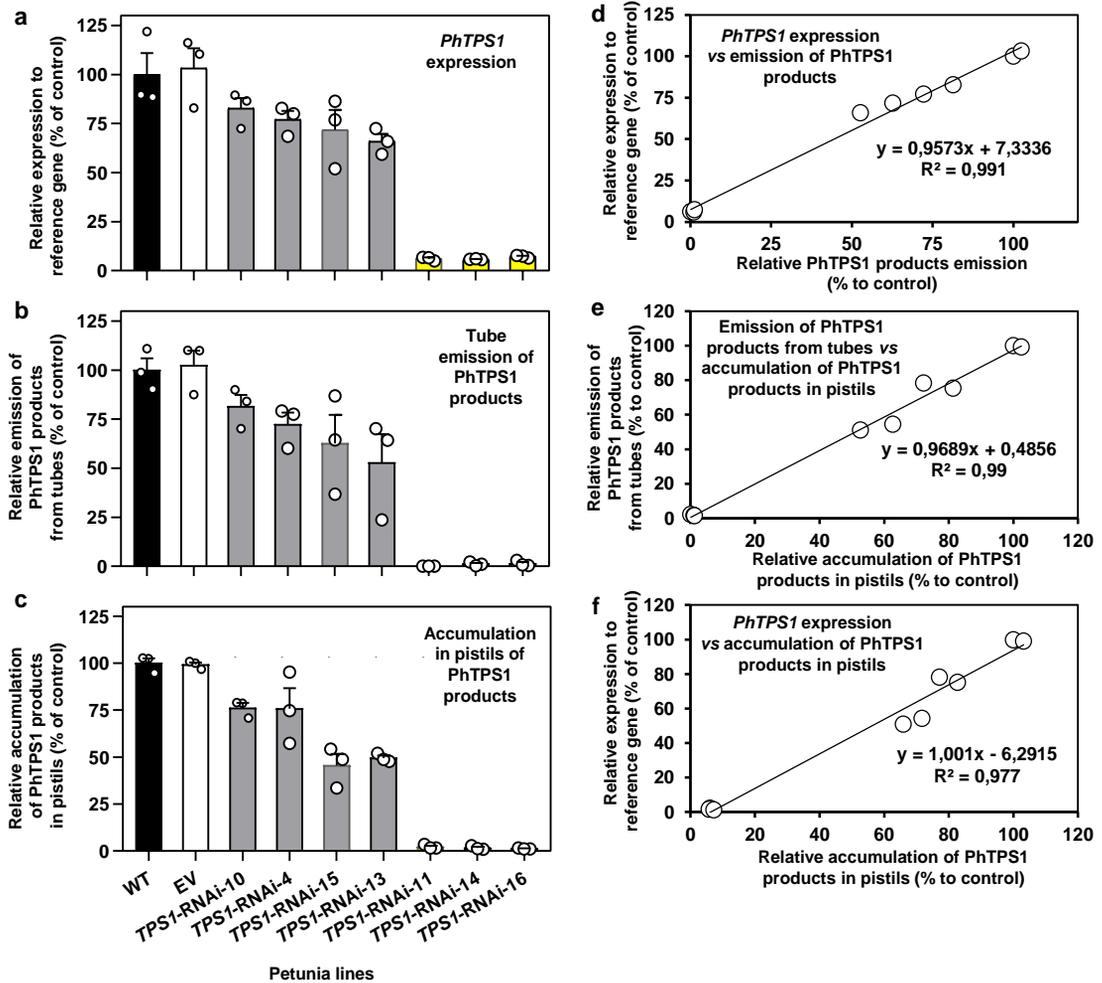

**Supplementary Fig. 11 | Screening of *PhTPS1*-RNAi lines and relationships between *PhTPS1* expression in tube, emission of volatiles from tube and internal pools in pistil.**
(**a**) *PhTPS1* mRNA levels on day 0 before anthesis in wild type (WT), an empty vector (EV) control and 7 independent *PhTPS1*-RNAi lines determined by qRT-PCR and shown relative to WT levels set as 100%. (**b**) Total emission rate of TPS1 volatile products from tubes and (**Cc**) their internal pools in pistils of WT, EV, and *PhTPS1*-RNAi flowers relative to the corresponding WT levels, set as 100%. Volatile emission from tubes was measured for 24 h starting at day 0 before anthesis. Internal pools from pistil were extracted on day 1 post-anthesis. Data in A-C are means ± SE (n = 3 biological replicates). (**d**) Correlation between *PhTPS1* expression and VOC emission from tubes. (**e**) Correlation between sesquiterpene emission from tubes and their internal pools in pistils. (**f**) Correlation between *PhTPS1* expression in tubes and internal pools of TPS1 products in pistils. RNAi lines used for further detailed analysis are shown in yellow.

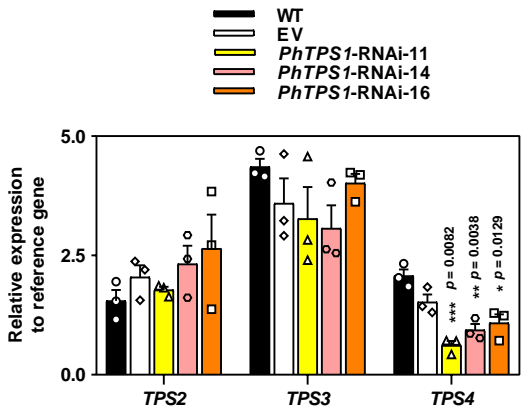

**Supplementary Fig. 12 | Effect of *PhTPS1*-RNAi downregulation on *PhTPS2*, *PhTPS3* and *PhTPS4* expression.**
*PhTPS2*, *PhTPS3* and *PhTPS4* mRNA levels on day 0 before anthesis in pistils from wild type (WT), an empty vector (EV) control and three independent *PhTPS1*-RNAi determined by qRT-PCR. *PP2A* was used as a reference gene. Data are means ± SE (n = 3 biological replicates). Significant *p* values ($p < 0.05$), as determined by two-tailed paired *t*-test relative to control, are shown.

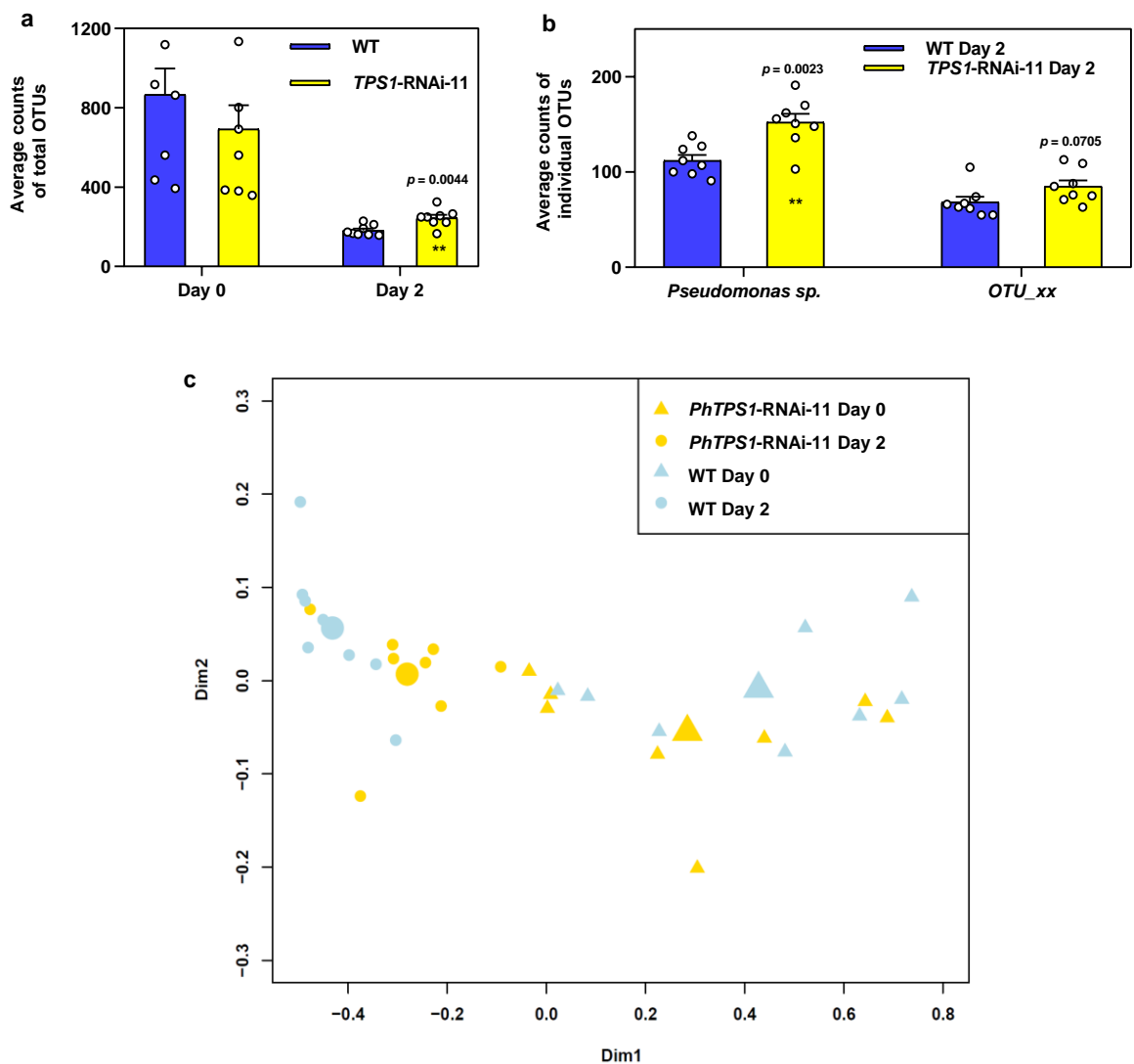

**Supplementary Fig. 13 | Effect of sesquiterpene fumigation on the microbiome community of pistil.**
(**a**) Average counts of total bacteria from 7 Operational Taxonomic Units (OTUs) in stigma microbiome of wild type (WT) and *PhTPS1*-RNAi flowers on day 0 before anthesis and day 2 after anthesis. Microbial DNA was extracted from the surface of WT and *PhTPS1*-RNAi-11 pistils and microbiome was profiled by sequencing. (**b**) Average counts of the 2 OTUs present on the WT and *PhTPS1*-RNAi-11 pistil surfaces on day 2 post-anthesis. In **a** and **b**, data are means ± SE (n = 8 biological replicates). Significant *p* values (p < 0.05), as determined by two-tailed Student t-test relative to control, are shown in **a** and **b**. (**c**) Non-metric multidimensional scaling (NMDS) based on Bray-Curtis Distances in community composition across the samples. Samples are clearly separated by time after anthesis and plant genotype (wild type or transgenic, fitting of factor onto ordination: $r^2$ = 0.7565, *p* = 0.000999). Large symbols denote the centroids of the four sample groups.

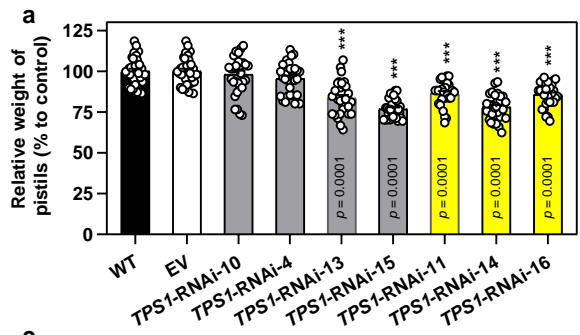
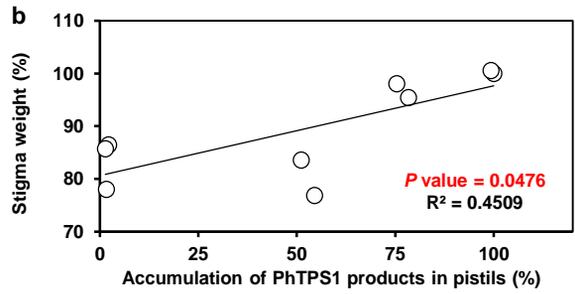
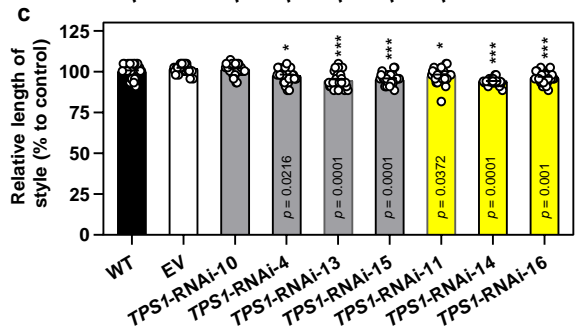
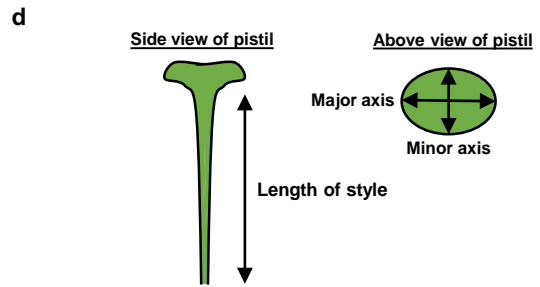
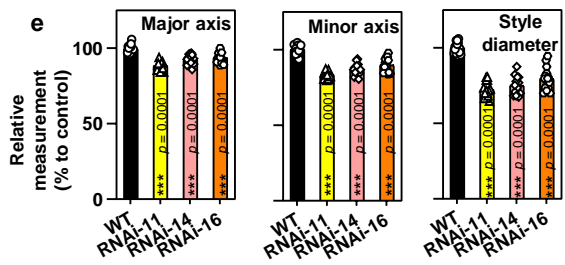
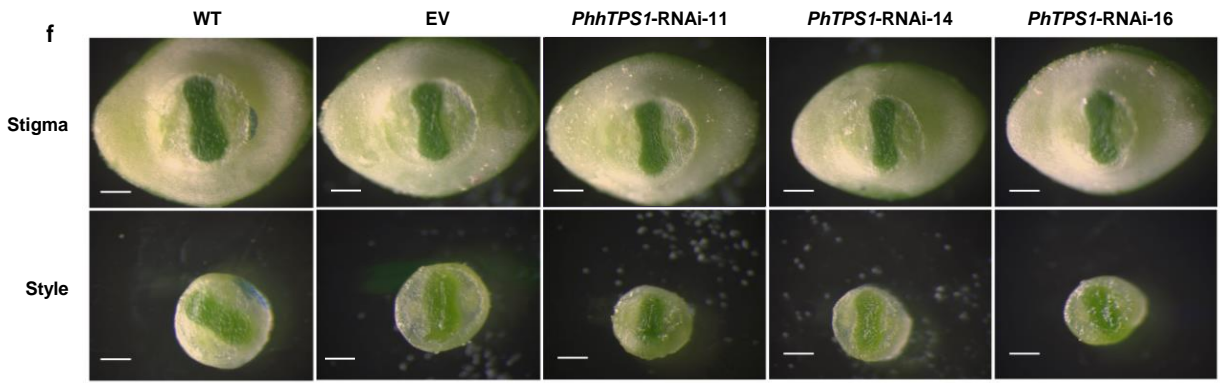
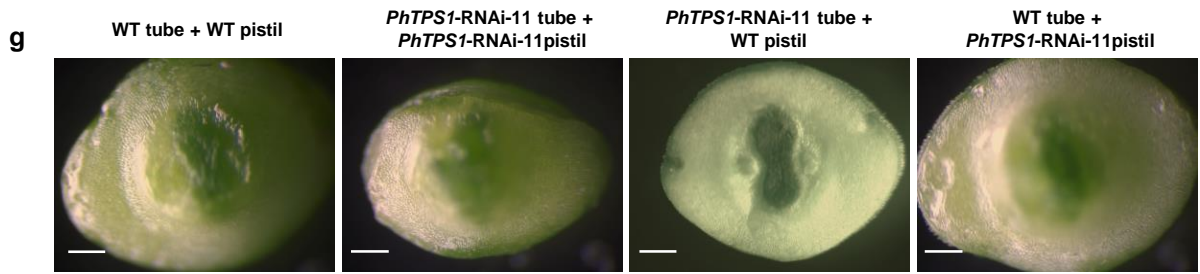

**Supplementary Fig. 14 | Effect of *PhTPS1* expression in tube on pistil development.**
Weight (**a**) and length (**c**) of pistil from wild type (WT), an empty vector (EV) control and *PhTPS1*-RNAi transgenic lines measured on day 1 post-anthesis relative to corresponding WT measurements set as 100%. Data are means ± SE (n = 30 biological replicates). (**b**) Correlation between stigma weight and accumulation of PhTPS1 products in stigmas of the different transgenic lines. (**d**) Scheme illustrating the dimension parameters of a petunia pistil. (**e**) Pistil parameters (as described in **d**) and style diameter on day 1 post-anthesis were measured under a fluorescence microscope. Data are means ± SE (n = 15 biological replicates). (**f**) Pictures of cross sections of stigmas and styles from wild type (WT), an empty vector (EV) control and *PhTPS1*-RNAi transgenic lines taken on day 1 post-anthesis. Pictures are representative of 15 independent observations for each line. (**g**) Pictures of cross sections of stigmas from the complementation experiments that generated results shown in **Fig. 3b.** Pictures were taken on day 1 post-anthesis. All scale bars = 300 µm. Pictures are in (**f**) and (**g**) are representative of 8 independent experiments for each line. Significant *p* values ($p < 0.05$), as determined by two-tailed paired *t*-test relative to control, are shown in (**a**), (**c**) and (**e**).

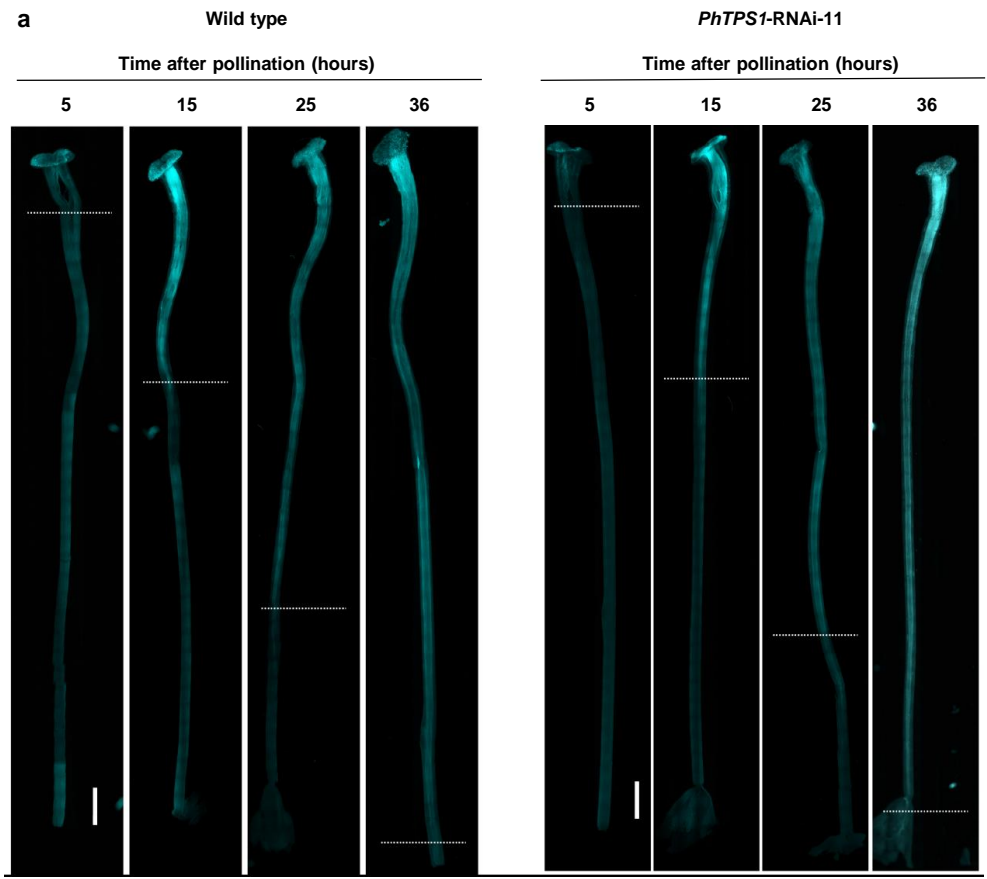

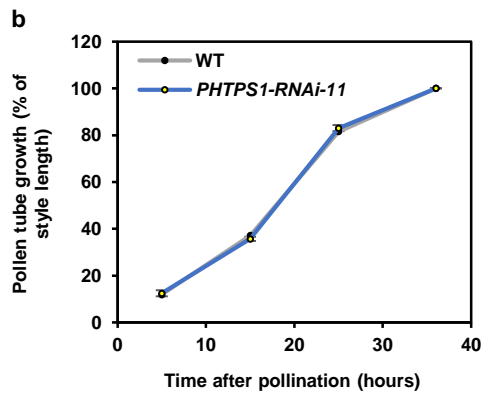

**Supplementary Fig. 15 | Effect of *PhTPS1* expression in tube on pollen germination and pollen tube growth through the style.**
(**a**) Pictures show the pollen tube growth at different times following pollination on representative WT and *PhTPS1*-RNAi-11 pistils. WT and *PhTPS1*-RNAi-11 petunia flowers were emasculated on day 1 post-anthesis and hand pollinated on day 2 post-anthesis at 18:00 h. At the indicated times, pistils were harvested and stained with aniline blue to measure pollen tube growth. Vertical bar = 2 mm. Horizontal dotted bars show the progression of the pollen tube growth. Pictures are representative of 4 independent experiments for each time point. (**b**) Pollen tube growth over time following pollination. Data are means ± SD (n = 4 biological replicates).

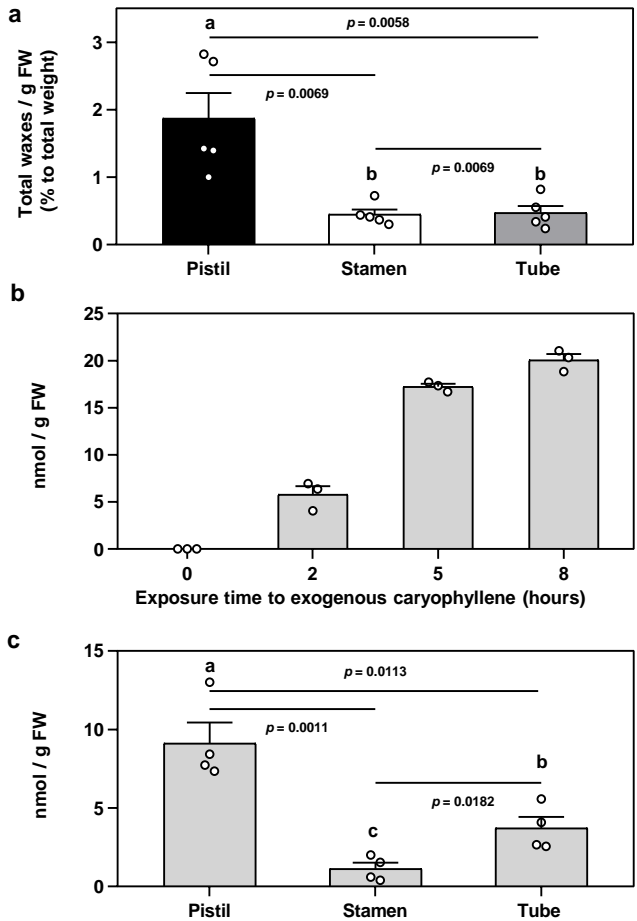

**Supplementary Fig. 16 | Quantification of cuticular waxes in petunia flower organs and quantification of exogenously fed caryophyllene.**
(**a**) Quantification of total waxes from petunia pistil, stamen and tube on day 0 before anthesis. Data are presented as total waxes as a proportion of the total fresh weight of each tissue, and are mean ± S.E. (n = 5 biological replicates). Different letters indicate statistically significant differences between tissues (ANOVA). (**b**) Caryophyllene content of petunia pistils from day 0 flowers incubated for indicated times in sealed vials containing 10 µmol of exogenous caryophyllene on a filter paper. Data are mean ± SE (n = 3 biological replicates). (**c**) Comparison of the accumulation of exogenously supplied caryophyllene by petunia pistil, tube, and stamen incubated for 5 h as performed in **b**. Data are mean ± SE (n = 3 biological replicates). Different letters indicate statistically significant differences between tissues (ANOVA).

**Table S1 | PCR primer list.**

| Primers used for "USER" cloning in yeast expression | |
|---|---|
| **Primer names and target genes** | **Primer sequences (5' -> 3')** |
| PhTPS1-CDS-FU | GGCTTAAUCGGGAGCAAACCTTACGAGA |
| PhTPS1-CDS-RU | GGTTTAAUAACAGGATCAGATACAACGTCA |
| PhTPS2-CDS-FU | GGCTTAAUCGATCTCCAACACTGCATGG |
| PhTPS2-CDS-RU | GGTTTAAUTAGAATTAGCACGTGGGGCC |
| PhTPS3-CDS-FU | GGCTTAAUGTGAAACTGCAAAGGAGGCT |
| PhTPS3-CDS-RU | GGTTTAAUCCACATATGCATTACGACGGT |
| PhTPS4-CDS-FU | GGCTTAAUATGAGTCAACCAGTTTGCTCCC |
| PhTPS4-CDS-RU | GGTTTAAUTCATACTTTGACAGTTTCGACAAG |

| Primers used for quantitative RT-PCR | |
|---|---|
| **Primer names and target genes** | **Primer sequences (5' -> 3')** |
| PhPP2A-qPCR-F | GACCGGAGCCAACTAGGAC |
| *PhPP2A-qPCR-R* | AAAACTTGGTAACTTTTCCAGCA |
| PhTPS1-qPCR-F | GCAACTGAAGCGCCTATGTT |
| PhTPS1-qPCR-R | TGTGTATCCATCCGCCTCTT |
| PhTPS2-qPCR-F | GGCATTGGAATCCTTACGCA |
| PhTPS2-qPCR-R | TGTGTTCTCTTGCCTCGTCT |
| PhTPS3-qPCR-F | GAGTGGGTGACAAATGAGCC |
| PhTPS3-qPCR-R | GACGTATGCCTCTTCCTCTGA |
| PhTPS4-qPCR-F | GCAAGAGGCGTACATTGAGC |
| PhTPS4-qPCR-R | CCATCACGCGAGCAAGATTT |

**Captions for Dataset 1.**

Microbiome analysis of flowers of *Brassica rapa* plants cultivated under sterile conditions from surface sterilized seeds (sample_33) or grown in soil in the lab after treatment with different bacteria. Microbiome analysis of pistils of wild type and *PhTPS1*-RNAi-11 *Petunia x hybrida* flowers.